\documentclass{aa}  

\usepackage{amsmath,amssymb,float}
\usepackage{graphicx}
\usepackage{xcolor}
\usepackage{multirow}
\usepackage{soul}
\usepackage[utf8]{inputenc}
\usepackage{txfonts,textcomp}
\usepackage{comment}

\usepackage[normalem]{ulem}

\begin{document}

\title{Investigating supernova-remnant X-ray maps with dendrogram-based techniques}

   \author{
   S. Calderaro\inst{1}
   \and
   G. Parisi\inst{1}
   \and
   V. Sapienza\inst{2,1}
   \and
   E. Greco\inst{2}
   \and
   G. Lo Bosco\inst{3}
   \and
   R. N. Mantegna\inst{1, 4}
   \and
   G. Marsella\inst{1}
   \and
   S. Miccich\`e\inst{1}
   \and
   S. Orlando\inst{2}
   \and
   M. Miceli\inst{1,2}
   \fnmsep
          }

   \institute{Dipartimento di Fisica e Chimica - Emilio Segr\`e, Universit\`a degli Studi di Palermo, Viale delle Scienze, Ed. 18, 90128, Palermo, Italy
   \and
   INAF- Osservatorio Astronomico di Palermo G. S. Vaiana, Piazza del Parlamento, 1, 90134, Palermo, Italy
      \and
   Dipartimento di Matematica ed Informatica, Universit\`a degli Studi di Palermo, Via Archirafi, 34, 90121, Palermo, Italy
      \and
Complexity Science Hub, Metternichgasse 8, 1030 Vienna, Austria
             }

   \date{Received; accepted}

  \abstract
  {X-ray images of supernova remnants provide important information regarding the spatial distribution of the physical and chemical parameters of the shocked plasma.}
   {Our objective is to provide a comprehensive and quantitative assessment of the similarities observed across narrow-band images and equivalent-width maps of a given supernova remnant. We also aim to perform comparative analyses between distinct remnants.}
   {We adopted two dendrogram-based techniques, namely a multivariate analysis and a deep-learning, graph-based approach (by exploring different metrics) for the study of Chandra X-ray maps of Cassiopeia A, G292.0+1.8, Tycho's SNR, Kepler's SNR, and W49B. We also tested the reliability of our results on synthetic X-ray maps produced from a 3D hydrodynamic simulation of Cassiopeia A.}
   {We identify clusters of narrow-band images (and equivalent-width maps) corresponding to emission lines of chemical elements that are expected to be synthesized in the same layers of the exploding star and/or that originate from plasma at similar temperatures. By comparing different objects, we clearly observe strong similarities among Type Ia remnants (namely Tycho's SNR and Kepler's SNR) and core-collapse remnants (Cassiopeia A and G292.0+1.8), with W49B being more similar to the latter objects. }
   {The methodology developed in this work provides a robust tool to measure the analogies and differences between narrow-band images and equivalent-width maps of a supernova remnant, offering clues on the distribution of the different chemical species and on the plasma temperature. Although tested on a limited sample of objects, the adoption of this tool for the classification of supernova remnants (Type Ia vs. core-collapse) yields promising results.}


   \maketitle

\section{Introduction} \label{intro}

Supernova remnants (SNRs) are major drivers of the chemical evolution of our Galaxy, enriching the interstellar medium with both intermediate-mass elements, which are mainly produced in core-collapse supernovae (SNe), and Fe-peak elements, mainly originating in Type Ia SNe. 
The supersonic expansion of the ejecta caused by the SN explosion drives powerful shocks, producing a hot shell of shocked ambient medium. The deceleration of this shell induces the formation of a reverse shock, which propagates through the ejecta themselves, heating them up to X-ray-emitting temperatures.

The spatial distribution of the physical and chemical properties of the ejecta in a SNR is the result of the mechanisms governing the parent SN explosion, and it carries information concerning the physics of the explosion itself. 
While in middle-aged SNRs the interaction of the ejecta with an inhomogeneous environment typically alters the morphological and kinematic properties of the ejecta, young SNRs are expected to ``keep a memory"  of the explosion properties.
The X-ray emission of young SNRs is typically ejecta-dominated, thus providing a powerful diagnostic tool to address the imprint of the SN on the remnant. 

X-ray maps of SNRs in selected narrow bands (centered at the energies of specific metal emission lines), together with equivalent-width (EW) maps have been extensively used to trace the distribution of the chemical composition of the ejecta, both in Type Ia and in core-collapse SNRs (e.g., \citealt{2004A&A...414..545C,2006A&A...453..567M,2015ApJ...805..120M,2019MNRAS.489.4444B,2019ApJ...872...31B,2020ApJ...893...90S}). 
Indeed, the spatial distribution of the chemical composition of the ejecta provides information on the stellar layers where the elements were forged, as well as the presence of anisotropies in the explosion process, which may cause, for example, local (or even global) inversions of the classical element stratification pattern (where heavier elements are more concentrated toward the inner part of the remnant). 
On the other hand, the comparison between narrow-band images and EW maps in different bands is typically qualitative and not supported by quantitative assessments of the similarities/differences among different images.

Morphological studies of SNRs through statistical X-ray image-analysis techniques have provided promising results. For example, through multipole expansion (power-ratio method) and wavelet-transform analysis, a methodology capable of distinguishing SNRs originating from core-collapse explosions from those of Type Ia explosions was obtained \citep{2011ApJ...732..114L}. Gaussian-mixture decomposition techniques were also adopted to try to spatially separate the emission associated with the ejected stellar fragments from that of the circumstellar medium hit by the shock wave \citep{2013ApJ...764...63B}.
However, several issues remain, the most striking case being the SNR W49B. While the strong deviations from circularity of its X-ray emission clearly point toward a core-collapse origin \citep{2011ApJ...732..114L}, both the lack of a compact remnant and the abundance pattern indicate a Type Ia SN origin \citep{2018A&A...615A.150Z}. The latter scenario is confirmed by the analysis of X-ray data through smoothed-particle-inference techniques \citep{2020ApJ...904..175S} and by hydrodynamic simulations supporting a spherically symmetric explosion (typical of Type Ia SNe) in a highly inhomogeneous environment \citep{2011MNRAS.415..244Z}. Nevertheless, indications of a collimated explosion were also shown (e.g., \citealt{2006A&A...453..567M,2013ApJ...764...50L}) and recently confirmed by the detection of bipolar ejecta flows \citep{2025ApJ...988L..58X}.
We stress that the classification method proposed by \citet{2011ApJ...732..114L} is merely based on the presence of deviations from spherical symmetry in the soft X-ray maps of the remnants and does not provide a metric to quantify the similarities among different SNRs. Moreover, the physical origin of the departures from spherical symmetry might be intrinsic (i.e., associated with anisotropies in the explosion), or extrinsic (due to inhomogeneities in the ambient medium).

To address these limitations, we present a novel approach by quantitatively measuring the similarities among different narrow-band (and EW) maps of a given SNR and of different SNRs. We also investigated the morphogenesis of the features observed in the X-ray maps by taking advantage of synthetic images derived from the three-dimensional (3D) hydrodynamic (HD) simulations of Cassiopeia A presented in \citet{2022A&A...666A...2O}. 
The methodology is based on the introduction of a measure of similarity among the SNR maps. This similarity can be computed by leveraging either standard multivariate analysis metrics or deep-learning ones. After similarities were computed, standard hierarchical clustering techniques were applied (e.g., \citealt{Anderberg1973ClusterApp}), thus revealing the hierarchical clustering of elements in the SNRs maps.
We tested our tool on \emph{Chandra} archive observations of Cassiopeia A, Tycho's SNR, Kepler's SNR, W49B, and G292.0+01.8, and on the synthetic X-ray images derived from the 3D HD simulations of Cassiopeia A. We identified clusters of different chemical elements in Type Ia and core-collapse remnants and obtained interesting clues to the origin of W49B.
To improve the solidity of our results and to allow for cross-validation, we approached our task from two different perspectives: the more classical one of clustering and network theory techniques, and the more modern one of artificial intelligence techniques. 

The paper is organized as follows. In Section~\ref{data} we report on the considered data and the pre-processing procedure. In Section~\ref{methods} we illustrate the similarity measures considered in our work and the dendrogram-based procedure used to classify SNRs. Our results are shown in Section~\ref{results}. We discuss our results and draw our conclusions in Section \ref{concl}.

\section{Data} \label{data}

\subsection{Observations and dataset creation}

\begin{table*}[t!]
    \centering
    \caption{\textit{Chandra}/ACIS observations table.}
    \begin{tabular}{cccccc}
    \hline\hline
    SNR & Obs ID & Exp. Time (ks) & R. A.                  & Dec.                      & Start Date \\
    \hline
    \multirow{8}{6em}{Cassiopeia A} & 4634 & 148.6 & 23$^h$ 23$^m$ 26.7$^s$ & 58\textdegree 49' 03.0'' & 28/04/2004 \\
    & 4635 & 135.0 & 23$^h$ 23$^m$ 26.7$^s$ & 58\textdegree 49' 03.0'' & 01/05/2004 \\
    & 4636 & 143.5 & 23$^h$ 23$^m$ 26.7$^s$ & 58\textdegree 49' 03.0'' & 20/04/2004 \\
    & 4637 & 163.5 & 23$^h$ 23$^m$ 26.7$^s$ & 58\textdegree 49' 03.0'' & 22/04/2004 \\
    & 4638 & 164.5 & 23$^h$ 23$^m$ 26.7$^s$ & 58\textdegree 49' 03.0'' & 14/04/2004 \\
    & 4639 & 79.0  & 23$^h$ 23$^m$ 26.7$^s$ & 58\textdegree 49' 03.0'' & 25/04/2004 \\
    & 5196 & 49.5  & 23$^h$ 23$^m$ 26.7$^s$ & 58\textdegree 49' 03.0'' & 08/02/2004 \\
    & 5319 & 42.3  & 23$^h$ 23$^m$ 26.7$^s$ & 58\textdegree 49' 03.0'' & 18/04/2004 \\
    & 5320 & 54.4  & 23$^h$ 23$^m$ 26.7$^s$ & 58\textdegree 49' 03.0'' & 05/05/2004 \\
    \hline
    \multirow{6}{6em}{Kepler's SNR} & 6714 & 157.8 & 17$^h$ 30$^m$ 42.0$^s$ & -21\textdegree 29' 00.0'' & 27/04/2006 \\
    & 6715 & 159.1 & 17$^h$ 30$^m$ 41.2$^s$ & -21\textdegree 29' 31.4'' & 03/08/2006 \\
    & 6716 & 158.0 & 17$^h$ 30$^m$ 42.0$^s$ & -21\textdegree 29' 00.0'' & 05/05/2006 \\
    & 6717 & 106.8 & 17$^h$ 30$^m$ 41.2$^s$ & -21\textdegree 29' 31.4'' & 13/07/2006 \\
    & 6718 & 107.8 & 17$^h$ 30$^m$ 41.2$^s$ & -21\textdegree 29' 31.4'' & 21/07/2006 \\
    & 7366 & 51.5  & 17$^h$ 30$^m$ 41.2$^s$ & -21\textdegree 29' 31.4'' & 16/07/2006 \\
    \hline
    \multirow{6}{6em}{G292.0+1.8} & 6677 & 159.1 & 11$^h$ 24$^m$ 39.1$^s$ & -59\textdegree 16' 20.0'' & 16/10/2006 \\
    & 6678 & 44.3 & 11$^h$ 24$^m$ 39.1$^s$ & -59\textdegree 16' 20.0'' & 02/10/2006 \\
    & 6679 & 153.9 & 11$^h$ 24$^m$ 39.1$^s$ & -59\textdegree 16' 20.0'' & 03/10/2006 \\
    & 6680 & 39.3 & 11$^h$ 24$^m$ 39.1$^s$ & -59\textdegree 16' 20.0'' & 13/9/2006 \\
    & 8221 & 65.0 & 11$^h$ 24$^m$ 39.1$^s$ & -59\textdegree 16' 20.0'' & 20/10/2006 \\
    & 8447 & 47.7 & 11$^h$ 24$^m$ 39.1$^s$ & -59\textdegree 16' 20.0'' & 07/10/2006 \\
    \hline
    \multirow{9}{6em}{Tycho's SNR} & 10093 & 118.4 & 00$^h$ 25$^m$ 19.0$^s$ & 64\textdegree 08' 10.0'' & 13/04/2009 \\
    & 10094 & 90.0 & 00$^h$ 25$^m$ 19.0$^s$ & 64\textdegree 08' 10.0'' & 18/04/2009 \\
    & 10095 & 173.4 & 00$^h$ 25$^m$ 19.0$^s$ & 64\textdegree 08' 10.0'' & 23/04/2009 \\
    & 10096 & 105.7 & 00$^h$ 25$^m$ 19.0$^s$ & 64\textdegree 08' 10.0'' & 27/04/2009 \\
    & 10097 & 107.4 & 00$^h$ 25$^m$ 19.0$^s$ & 64\textdegree 08' 10.0'' & 11/04/2009 \\
    & 10902 & 39.5 & 00$^h$ 25$^m$ 19.0$^s$ & 64\textdegree 08' 10.0'' & 15/04/2009 \\
    & 10903 & 23.9 & 00$^h$ 25$^m$ 19.0$^s$ & 64\textdegree 08' 10.0'' & 17/04/2009 \\
    & 10904 & 34.7 & 00$^h$ 25$^m$ 19.0$^s$ & 64\textdegree 08' 10.0'' & 13/04/2009 \\
    & 10906 & 41.1 & 00$^h$ 25$^m$ 19.0$^s$ & 64\textdegree 08' 10.0'' & 15/04/2009 \\
    \hline
    \multirow{2}{6em}{W49B} & 13440 & 157.6 & 19$^h$ 11$^m$ 06.6$^s$ & 09\textdegree 06' 00.0'' & 18/08/2011 \\
    & 13441 & 59.7 & 19$^h$ 11$^m$ 06.6$^s$ & 09\textdegree 06' 00.0'' & 21/08/2011 \\
    \hline
    \end{tabular}
    \label{tab:obs}
\end{table*}
For our analysis, we made use of several archival \textit{Chandra}/ACIS observations of five different SNRs, as summarized in Table~\ref{tab:obs}.
Among the considered sources, two objects are classified as core-collapse SNRs (Cassiopeia A and G292.0+1.8) and two as Type~Ia SNRs (Tycho's and Kepler's SNR), while the origin of W49B is still debated in the literature, as explained in Sect. 1.
These targets were chosen primarily for their total exposure times, which provide high-quality photon statistics suitable for imaging.
Data were analyzed with \textit{CIAO} (version 4.15) using \textit{CALDB} (version 4.10.7).
All observations were reprocessed with the \texttt{chandra\_repro} task. 
For the image analysis, we used the \texttt{merge\_obs} command to generate fluxed, exposure-corrected, merged images and merged event files.

For the construction of the dataset, for each remnant we created a set of images slicing the spectrum into ten bands, each including specific ion emission lines---or continuum emission---as specified in Table \ref{tab:bands}.
We binned the images to an effective pixel size of 2 arcsec (4 bins of the ACIS detector) to ensure sufficient signal in all energy bands, without significantly degrading the spatial resolution.

\begin{table}[]
    \centering
    \caption{Energy bands selected with their corresponding emission lines and labels used throughout this work. }
    \begin{tabular}{ccc}
    \hline\hline
         Emission line & Label & Energy Band \\ \hline
         Fe L &Fe L& 0.5 - 1.2 keV\\
         Mg XI \& Mg XII &Mg tot& 1.2 - 1.6 keV\\
         Si XIII \& Si XIV &Si tot& 1.6 - 2.1 keV\\
         S XV \& S XVI &S tot& 2.1 - 3.0 keV\\
         Ar XVII \& Ar XVIII &Ar tot& 3.0 - 3.5 keV\\
         Ca XIX \& Ca XX &Ca tot& 3.5 - 4.0 keV\\
         Continuum 1&Cont 1& 4.0 - 5.0 keV\\
         Continuum 2&Cont 2& 5.0 - 6.0 keV\\
         Fe XXV&Fe XXV& 6.0 - 7.0 keV\\
         Ni XXVI/Continuum 3&Ni XXVI/Cont 3& 7.0 - 8.0 keV\\
         \hline
    \end{tabular}
    \label{tab:bands}
\end{table}
Besides this, we produced EW maps in a similar manner to \cite{2006A&A...453..567M}.
They are defined as 
\begin{equation}
    EW = \frac{F_{line}-F_{cont}}{F_{cont}}, \label{EWmapsdef}
\end{equation}
where $F_{line}$ is the flux due to the line at a given energy band plus the continuum emission, and $F_{cont}$ is the flux due to the continuum emission in that energy band. EW maps depend linearly on the abundances, but they are also affected by the temperature, column density, and ionization parameter.
The continuum beneath each emission line was estimated as follows. We extracted the global spectrum of the remnant and fitted the continuum in the $4.0-5.0$ keV band (which is free of strong line features) with an absorbed power law. We then adopted the WebPIMMS tool\footnote{\url{https://heasarc.gsfc.nasa.gov/cgi-bin/Tools/w3pimms/w3pimms.pl}} to extrapolate the continuum flux in other bands. We thus derived a multiplicative factor to scale the map in the 4.0--5.0 keV band to the desired energy band, thereby providing an image that represents the continuum emission underlying each emission line.

\subsection{Synthetic observations from the 3D model}

As a further test to validate the algorithms presented in this paper, we also synthesized the thermal X-ray emission from a 3D HD simulation describing Cassiopeia A \citep{2022A&A...666A...2O}. For the synthesis, we employed an in-house developed tool in a similar fashion to \cite{2019NatAs...3..236M}, \cite{2020A&A...638A.101G}, \cite{2021A&A...649A..14U}, and \cite{2024ApJ...961L...9S}. For each computational cell, we derived the physical quantities that govern the X-ray emission. In particular, we computed the electron number density, $n_{\rm e,j}$; the ion number density, $n_{\rm Z,j}$; and the cell volume, $V_{\rm j}$, from which the emission measure is defined as $\mathrm{EM}_{\rm j} = n_{\rm e,j} n_{\rm Z,j} V_{\rm j}$. The ionization state of the plasma is characterized by the ionization age $\tau_{\rm j} = n_{\rm e,j} \Delta t_{\rm j}$, where $\Delta t_{\rm j}$ represents the time elapsed since the plasma in the cell was shocked. The electron temperature evolution was treated by accounting for deviations from nonequilibrium electron–ion thermalization (see \citealt{2015ApJ...810..168O} for the details of the implementation). At the shock front, electrons are initialized at $kT = 0.3$ keV, consistently with instantaneous heating by lower hybrid waves (\citealt{2007ApJ...654L..69G}). Downstream, the electron temperature, $T_{\rm e,j}$, evolves through Coulomb collisions with ions and is computed as a function of the local plasma conditions and $\Delta t_{\rm j}$. 

These quantities ($T_{\rm e,j}$, $n_{\rm e,j}$, and $\tau_{\rm j}$), together with the local chemical composition, are used as input parameters for the nonequilibrium ionization, optically thin plasma emission model \texttt{vnei} implemented in XSPEC (version 12.14.1, ATOMDB version 3.0.9; \citealt{1996ASPC..101...17A}). The chemical abundances in each cell were not assumed a priori but were self-consistently determined through passive tracers embedded in the simulation (see \citealt{2025A&A...699A.305O} for details of the implementation). These tracers follow the spatial and temporal evolution of the elements synthesized both during the progenitor’s stellar evolution and in the SN explosion itself. For each cell, we computed the emitted X-ray spectrum in selected energy bands and integrated the emission along the line of sight to obtain synthetic flux maps. In analogy with the analysis of \textit{Chandra} observations, we also produced images and EW maps in the same energy bands.

This methodology is particularly powerful because it preserves the coupling between dynamics, thermodynamics, and chemical stratification. As a result, it enables a direct link between the spatial distribution of elements in the remnant and the physical processes that shaped the explosion. In particular, the inclusion of tracer-based abundances allows us to associate observed asymmetries and compositional inhomogeneities in the remnant with anisotropies in the SN mechanism and with the HD evolution of the ejecta. This provides a robust framework for testing the capability of the dendrogram-based methodology to recover information about the SN explosion from SNR observations while offering a physically grounded interpretation of the observed X-ray morphology.

\subsection{ Data pre-processing} \label{pulizia}

For a proper comparison among the different images, we removed all pixels with a value of zero (in both the flux images and the EW maps) and resized all images to the same dimensions.
The data pre-processing pipeline consists of three steps to address the issues mentioned above. First,  we reduced the size of each image to that of the smallest rectangle including the SNR (defined by the pixels whose flux in the $0.3-8$ keV band is 5 times larger than that of the background) that contains all nonzero pixels. Secondly, we converted the cropped image into a square by adding zero-padding. The padding was applied symmetrically along each axis, with an additional pixel added when necessary to handle odd–even differences between the image size and the target dimension. This ensures that each cropped image is transformed into a square whose side length is equal to the largest dimension of the original cropped region. In the last step, we rescaled all images to ($256 \times 256$) pixels using the \texttt{resize} function from the Python package \texttt{PILLOW}.

This resizing step employs the Lanczos filter \citep{1964SJNA....1...76L}, a high-quality resampling filter based on a windowed sinc function. The Lanczos filter is particularly effective for downsampling as it minimizes aliasing artifacts and preserves image sharpness and detail more effectively than simpler interpolation methods such as nearest-neighbor or bilinear interpolation. By using the Lanczos filter, the pipeline ensures that the resized images maintain high fidelity, which is crucial for subsequent analysis and comparison. This uniform sizing is fundamental for our experimental activity since it allows us to perform comparisons not only between elements of the same SNR but also across different SNRs.

To test the impact of each preprocessing step on the results, we repeated the analysis described above starting from the original images and progressively applied the successive procedures. The presence of background may affect the dendrogram structure; for this reason a region of interest (ROI) was selected for each image to minimize background influence. The image was cropped according to the size of the bounding rectangle that contains the ROI. Neither the padding nor the resizing procedures produced appreciable changes. We further evaluated two interpolation methods using the Lanczos and nearest-neighbor filters. Although minor differences were occasionally observed in the dendrogram heights at which some elements grouped, the overall hierarchical organization and cluster assignments also remained highly consistent across interpolation methods.

\section{Methods}  \label{methods}

In this section we illustrate the different methodologies used to quantify the similarity amongst SNRs and the clustering procedures able to investigate their hierarchical organization in clusters.

\subsection{Multivariate-analysis similarity measures}  \label{corr}

Multivariate analysis deals with observations on $N>1$ variables that show some inherent interdependence amongst them. In the simplest case the $i$th variable is represented by a numerical array or vector, $V_i=\{ v_{i,1}, v_{i,2}, \cdots v_{i,T}\},$ of length $T$. The interdependence is measured by introducing a similarity measure among pairs of such vectors. Typical similarity measures are the Pearson correlation coefficient, defined as
\begin{eqnarray}
    \rho_{ij}={{\langle V_i V_j \rangle - \langle V_i \rangle \langle V_j \rangle} 
    \over 
    {\sqrt{\langle |V_i -\langle V_i \rangle |^2 \rangle} \, \sqrt{ \langle |V_j - \langle V_j \rangle|^2\rangle}}}
    \qquad 
    \langle V_i \rangle = {1 \over T} \, \sum_{k=1}^T v_{i,k}
    \label{pearson}
,\end{eqnarray}
or the Cosine distance defined as
\begin{eqnarray}
    {d_{ij}^{cos}}=1-{\rho_{ij}^{cos}} \qquad \qquad 
    {\rho_{ij}^{cos}}=\frac{V_i \cdot V_j}{\sqrt{\|V_i \|^2} \, \sqrt{ \|V_j \|^2}} \label{cos}
,\end{eqnarray}
where $V_i \cdot V_j= T \langle V_i V_j \rangle$ denotes the usual scalar product and  $\| \cdot \| $ is the Euclidean $L_2$ norm. The resulting value lies in the $[0, 2]$ range, where 0 indicates identical vectors, 1 indicates no correlation, and 2 indicates different vectors. Cosine distance is a commonly used dissimilarity measure that quantifies the angular distance between two nonzero vectors in an inner product space. On the other hand, the Pearson correlation coefficient is a typical similarity measure whose values lie in the $[-1, +1]$ domain, with positive values indicating correlations amongst data, negative values indicating anticorrelation, and 0 indicating that the two vectors are independent of each other. Starting from $\rho_{ij}$, we constructed a dissimilarity measure or distance by simply considering $d_{ij}=\sqrt{2 \, (1 - \rho_{ij})}$.
It is worth mentioning that these are just two examples taken out of a plethora of similarity measures that can be considered depending on the nature of the variables one is dealing with.

When working with standardized variables with null average and unitary variance, we have $\rho_{ij}={\rho_{ij}^{cos}}$. In this case, the differences in using the two above measures are only observed at the level of distances. In fact, the cosine distance is a linear function of ${\rho_{ij}^{cos}}$, while the distance $d_{ij}=\sqrt{2 \, (1 - \rho_{ij})}$ introduces some nonlinearities that might amplify differences between $V_i$ and $V_j$. When dealing with non-standardized variables, the role of averages and variances is discounted by the Pearson correlation coefficient, $\rho_{ij}$, while they might play a relevant role in $\rho_{ij}^{cos}$.

In our case, however, data are bi-dimensional arrays. Specifically, each image relative to the $i$th considered energy band will be represented by an $n \times n$ matrix $M_i = \{ m_{kj}^{(i)}, j,k=1 \cdots n \}$, where $n$ is the number of aggregated pixels ($n=256$ in our case).

Also in this case there are several similarity measures or distances between matrix pairs.  In our study, we selected the two specific measures detailed below.
\begin{enumerate}
\item The normalized inner product of Frobenius is defined as
\begin{eqnarray}
   && \rho_{ij}={{tr \bigl(M_i \, M_j^t \bigl)}
    \over
    {|| M_i ||_F \, || M_j ||_F}}, \\
   && || M_i ||_F=\sqrt{\sum_{k=1}^n \sum_{j=1}^n (m_{kj}^{(i)})^2}=\sqrt{tr \bigl( M_i M_i^t\bigl)} 
    \label{frob}
,\end{eqnarray}
where the superscript $t$ stands for the {\em{\emph{transpose}}} matrix operation and the $tr$ symbol indicates the {\em{\emph{trace}}} matrix operation. It is worth mentioning that $tr \bigl( M_i M_j^t \bigl) =\sum_{r=1}^n \sum_{s=1}^n m_{rs}^{(i)} m_{rs}^{(j)}$, which indicates that the Frobenius inner product essentially involves the sum of the two matrix elements taken in the same position, which is similar to what is done for 1-D arrays. We then obtained a distance among matrices by using $d_{ij}=\sqrt{2 \, (1 - \rho_{ij})}$.
\item The cosine distance of Eq. \ref{cos} on matrices. Specifically, we first obtained a vector, $V_i$, from matrix $M_i$ by simply adding matrix rows one after the other, and then we used the cosine distance.
\end{enumerate}
The same considerations made for the Pearson correlation coefficient and the cosine distance between vectors apply to the Frobenius distance and cosine distance between matrices. We used both measures to investigate how robust our results are with respect to the role of averages, variances, and the nonlinearities introduced by the considered distance.

\subsection{Dendrogram-based techniques}  \label{HT}

A widespread unsupervised approach used to detect the information associated with the similarity matrix is given by the hierarchical clustering (HC) analysis.  HC is a popular data analysis procedure that groups elements of a set into a hierarchy of clusters \cite{HanKamberPei2011DMCT}. It is widely used in many research fields. Examples include computational biology  \cite{Felsenstein1981EvoTrees}, genomics \cite{Eisen1998ClusterGenomewide}, neuroscience \cite{Filzmoser1999HierarchicalfMRI, GOUTTE1999298, BAUNE1999477}, psychology  \cite{Edelbrock01071979}, finance  \cite{Mantegna1999HierarchicalStructure, Tumminello_2007, MUSCIOTTO2016267}, and economics  \cite{gligor2008convergence}, as well as social sciences  \cite{Newman2001, sales2007extracting}. 

If we consider a set of $n$ objects and suppose that a similarity measure (see Sect. \ref{corr}) between pairs of elements is defined, the HC methods allow us to hierarchically organize the elements in clusters. The result of the procedure is a rooted tree, also called a {\em{\emph{dendrogram}}}, that gives a quantitative description of the clusters thus obtained. 

A large number of HC procedures can be found in the literature \citep{Anderberg1973ClusterApp, coronnello2005sectoridentificationsetstock}. We adopted the average linkage cluster analysis (ALCA) here.

The ALCA is an HC procedure \citep{Anderberg1973ClusterApp} that can be described by considering either a similarity or a distance measure. Here we consider the distance matrix D. The following procedure performs the ALCA, giving a rooted tree and an ultrametric matrix $D^{<}$ of elements $d_{ij}^<$ as an output.
\begin{enumerate}
\item Set $T$ as the matrix of elements such that $T= D$.
\item Select the minimum distance, $t_{hk}$, in the distance matrix, $T$. Note that after the first step of construction, $h$ and $k$ can be simple elements (i.e., clusters of one element each) or clusters (sets of elements).
\item Merge cluster $h$ and cluster $k$ into a single cluster, say $h$. The merging operation identifies a node (also called a clade) in the rooted tree connecting clusters $h$ and $k$ at the distance $t_{hk}$. Furthermore, to obtain the ultrametric matrix, it is sufficient that for $\forall i \in h$ and $\forall j \in k$ one sets $d_{ij}^<=d_{ji}^<=t_{hk}$.
\item Redefine the matrix $T$:
\begin{eqnarray}
    \left \lbrace 
    \begin{aligned}  
    &  t_{hj}={{N_h t_{hj}+N_k t_{kj}}\over{N_h+N_k}} {\rm{\, if \, j\ne h \, {\rm{and}} \, j \ne k}} \\
    & \\
    &   t_{ij}=t_{ij} {\rm{\, \, otherwise}}
    \end{aligned}
    \right.                                          
,\end{eqnarray}
where $N_h$ and $N_k$ are the number of elements belonging, respectively, to the cluster $h$ and the cluster $k$. Note that if the dimension of $T$ is $m \times m$, then the dimension of the redefined $T$ is $(m-1) \times (m-1)$  because of the merging of clusters $h$ and $k$ into the cluster $h$.
\item If the dimension of $T$ is bigger than one, then return to step 2; otherwise, stop.
\end{enumerate}

The algorithm can be easily adapted for working with similarities instead of distances. It is just enough to exchange the distance matrix, D, with a similarity matrix (e.g., the correlation matrix) and replace the search for the minimum distance in the matrix T in point 2 of the above algorithm with the search for the maximum similarity. It is worth noting that the ALCA can produce different hierarchical trees depending on the use of a similarity matrix or a distance matrix.

Once a dissimilarity (or similarity) measure between elements is defined and a clustering procedure is selected, the HC algorithm is fully defined. The algorithm is deterministic and provides a hierarchical tree (or dendrogram) as an output. However, the generation of a dendrogram does not mean that one also obtains a parsimonious hierarchical nested partition as an output of the HC. Historically, the simplest and most popular way to obtain a partition from a hierarchical tree was to cut the dendrogram at a fixed dissimilarity (or similarity) value. With this simple approach, such a cut defines the composition of clusters. They are selected by considering the groups of elements linked in the tree at a dissimilarity value smaller than the threshold value. 

Operatively, a dendrogram is a rooted tree where the vertical lines (leaves) represent the elements of the systems under investigation. These leaves (or elements) are joined at different levels depending on their similarity/distance. The levels at which they join are the horizontal segments, also called clades. In the dendrograms we show hereafter, the quantities shown on the vertical axis are distances. Therefore, the lower the clade the shorter the distance, for example, the more similar the elements are.

In \cite{BONGIORNO2022126933}, the statistically validated hierarchical clustering (SVHC) algorithm is proposed; it is based on bootstrap resampling of the multivariate numerical series under consideration that associates a p value at each clade of a given hierarchical tree. By adopting this algorithm, we can check whether or not the elements and clusters linked at that clade value can be considered as belonging to the same cluster. In particular, clusters associated with statistically significant p values can be regarded as robust and reliable. In the corresponding ultrametric matrix, these statistically significant clusters are bordered in green, while they are colored with a color other than black in the dendrograms, so that the validated blocks are immediately identifiable.

In general, when displaying dendrograms,  the vertical lines, representing the elements, are always colored in black. A horizontal line, i.e., a clade, is colored whenever the vertical lines attached to it represent elements that belong to a statistically significant (sub)cluster. In particular, for each clade of the dendrogram, a color is assigned corresponding to the smallest statistically validated cluster that contains it. Clades that do not belong to any validated cluster other than the root cluster are colored black, otherwise a color is chosen randomly. This coloring scheme helps to identify the statistically significant groups across different levels of the dendrogram. In the corresponding ultrametric matrix, these statistically significant clusters are highlighted in green. When presenting dendrograms including different SNRs the labels of the elements are colored in the dendrograms. Different colors refer to different SNRs.

\subsection{Deep-learning graph-based approach}  \label{GNN}

Another approach developed for clustering the X-ray maps of SNRs is based on graph neural networks (GNNs). GNNs are neural networks designed to work with graph-structured data, where information is not only tied to individual objects (nodes) but also to the relationships between them (edges). In this framework, each node in the graph represents an available image of an SNR, while edges encode the similarity between images. The model updates the representation of each node by combining information from its neighboring nodes with its own features. This allows the network to capture both the characteristics of individual images and the overall structure of the graph. When experiments are conducted across multiple SNRs, all images are integrated into a single graph, and the GNN is employed to learn node representations (embeddings) that can then be used for clustering. Intuitively, a GNN works through the two main steps described below
\begin{itemize}
    \item AGGREGATE: each node collects information from its neighboring nodes, i.e., the nodes it is connected to. It can be compared to a person gathering advice or opinions from friends.
    \item COMBINE: the node updates its own representation by combining the aggregated information from neighbors with its own features. In other words, the node decides how much of the “advice from friends” to incorporate into its own knowledge.
\end{itemize}
The rationale behind these steps is to allow each node to improve its own understanding by taking into account the features of its neighbors while retaining its original information. By repeating this process across multiple layers, each node gradually integrates information from a larger portion of the graph, effectively “learning from its wider community” and capturing both local and global patterns.
This process is repeated for multiple layers, allowing each node to gradually incorporate information from a larger portion of the graph. For our implementation, we used a combination of two widely used architectures: graph convolutional networks (GCN) \citep{kipf2017semisupervised}, which update a node’s representation by averaging information from its neighbors, followed by a linear transformation and a nonlinear activation; and GraphSAGE \citep{hamilton2017inductive}, which generalizes GNNs by learning an aggregation function from neighbors rather than using a fixed average. For each node, a set of neighbors is sampled and their features are aggregated (e.g., by mean or max pooling) and combined with the node’s own features.
Before introducing the proposed GNN architecture, it is essential to describe the graph construction process. Starting from the set of SNR images, each image was first linearized into a 1D feature vector. We then computed a symmetric distance matrix, $D$, where each entry of $D_{ij}$ represents the pairwise cosine distance between the corresponding flattened images. 
To build the graph, we treated each SNR image as a node, while edges were established following a k-nearest-neighbor (Knn) strategy. Specifically, for each node, we connected it to its $k$ closest neighbors according to the cosine distance (see Eq.~\ref{cos}), ensuring a structured and meaningful representation of the similarity relationships among the images and obtaining a final graph of $\mathcal{G}=(\mathcal{V},\mathcal{E},X)$, where $\mathcal{V}$ is the set of nodes (the SNR images), $\mathcal{E}$ is the set of edges (in terms of image similarity), and finally $X \in \mathbb{R}^{\lvert \mathcal{V} \lvert \times C}$, where $C=256 \times 256$ is the intial node feature representation.
Essentially, $k$ controls the connectivity of the graph; higher  $k$ values lead to a denser graph structure, while lower values result in a sparser graph that captures only the most relevant similarity relationships.
The generated graph is used to train a GNN model capable of obtaining refined node embeddings. In particular, the GNN model we built consists of several convolutional layers that combine GraphSAGE and GCN, with batch normalization, residual connections, and dropout to mitigate overfitting. The model takes the node feature matrix, $X,$ and the graph connectivity, $\mathcal{E,}$ as input and produces a refined node embedding matrix of $Z \in \mathbb{R}^{\lvert  \mathcal{V} \rvert \times d}$, where $d$ is the dimensionality of the embedding space. To train the model in an unsupervised manner, we adopted a link-prediction objective.
Let $\mathcal{E}^+ = {(i,j) \in \mathcal{E}}$ the set of observed links and $\mathcal{E}^- \subseteq \mathcal{V} \times \mathcal {V}, \quad \mathcal{E}^- \cap \mathcal{E} = \emptyset$,  the set of nonexistent edges (negative samples). For each edge $(i,j)$, the model computes a score defined as $s_{ij} = z_i^\top z_j$ where $z_i, z_j \in \mathbb{R}^d$ are the final node embeddings. The negative edges in $\mathcal{E}^-$ are sampled dynamically during training; i.e., for each positive edge $(i,j) \in \mathcal{E}^+$, a set of nodes not connected to $i$ is randomly selected to form negative pairs $(i,k)$. This ensures that the model learns to distinguish connected from non-connected nodes while keeping the negative samples diverse and balanced relative to the positive edges. The loss used is a binary cross-entropy loss applied to the scores of positive and negative edges, which was designed to encourage the model to produce similar embeddings for connected nodes and dissimilar embeddings for non-connected nodes. Formally, this reads as
\begin{equation}
    \label{eq:lossgnn}
    \mathcal{L} = - \sum_{(i,j) \in \mathcal{E}^+} {\log {\sigma(s_{ij})}}  - \sum_{(i,j) \in \mathcal{E}^-}{\log(1-\sigma(s_{ij}))}
,\end{equation}
where $\sigma(\cdot)$ is the sigmoid activation function. The first term in the loss penalizes the model when connected nodes have low similarity, while the second term penalizes the model when non-connected nodes have high similarity. Minimizing $\mathcal{L}$ therefore encourages the embeddings, $Z,$ to respect the underlying graph structure. Nodes that are connected in $\mathcal{G}$ are mapped to nearby points in the embedding space, while unconnected nodes are mapped farther apart.
Once the network had been trained, the final node embeddings were used to perform HC using the average linkage method, allowing us to identify meaningful groups of similar nodes based on their learned representations.

We point out that the HC procedure used throughout the manuscript is always the same: i.e., average linkage. When considering the GNN approach or the cosine distance as a similarity measure, we considered the standard average linkage procedure. When considering the Frobenius distance, we used the SVHC, which is an augmented version of the average linkage procedure; here, when appropriate, the groups of elements that are joined together and potentially form a cluster are highlighted by colors randomly assigned to the clade. Therefore, the different approaches presented in the manuscript show the contribution of the similarity metric precisely.

\section{Results} \label{results}

In this section, we report on the results obtained by applying the methodologies discussed above to the preprocessed data mentioned in Section \ref{data}. We first report results on different images of a single SNR and then show the comparative analysis of multiple remnants.

\subsection{Single SNR investigation} \label{resultsfluxsingle}

\begin{figure}[htb] 
    \centering 
\includegraphics[width=0.7\columnwidth]{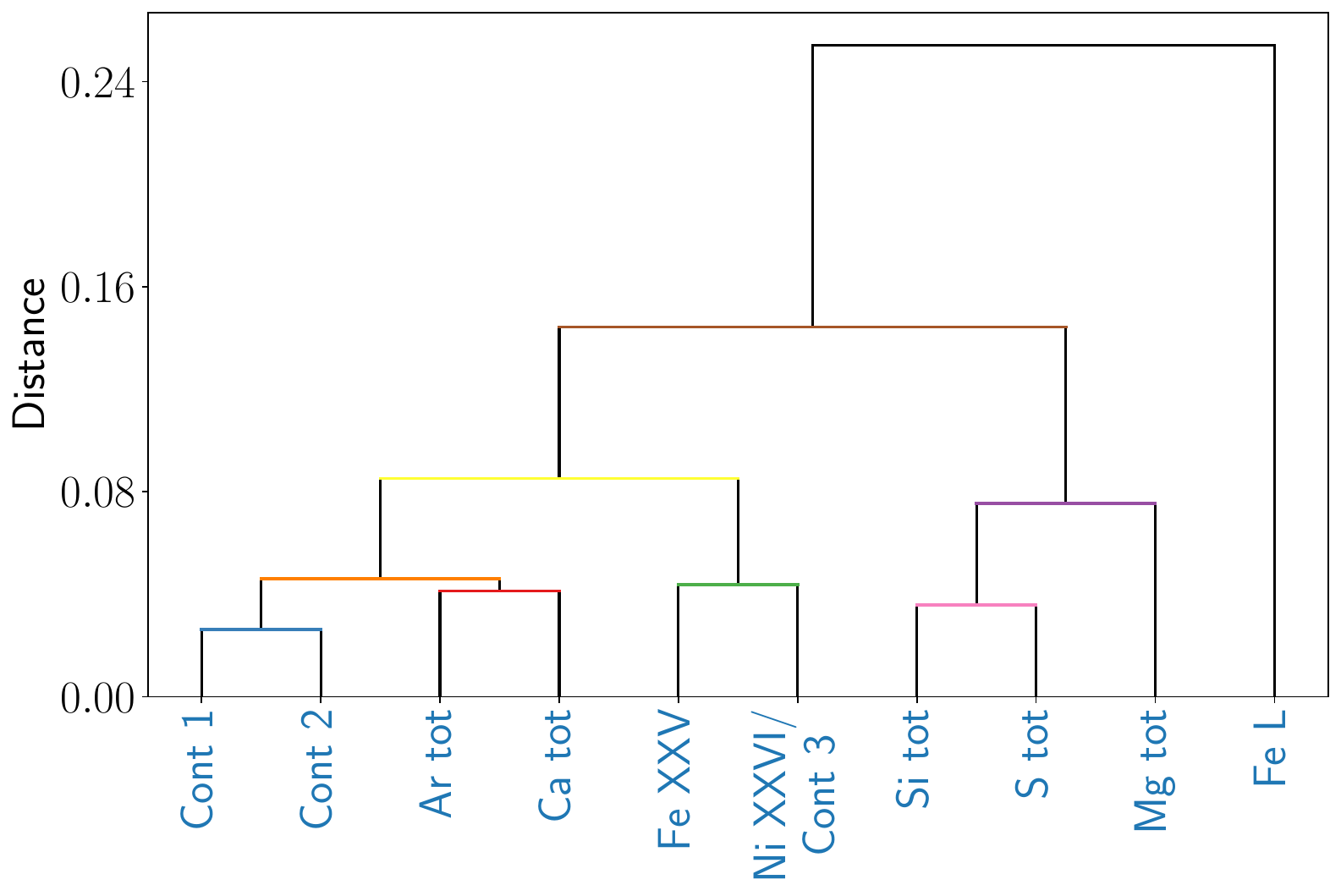}

\includegraphics[width=0.7\columnwidth]{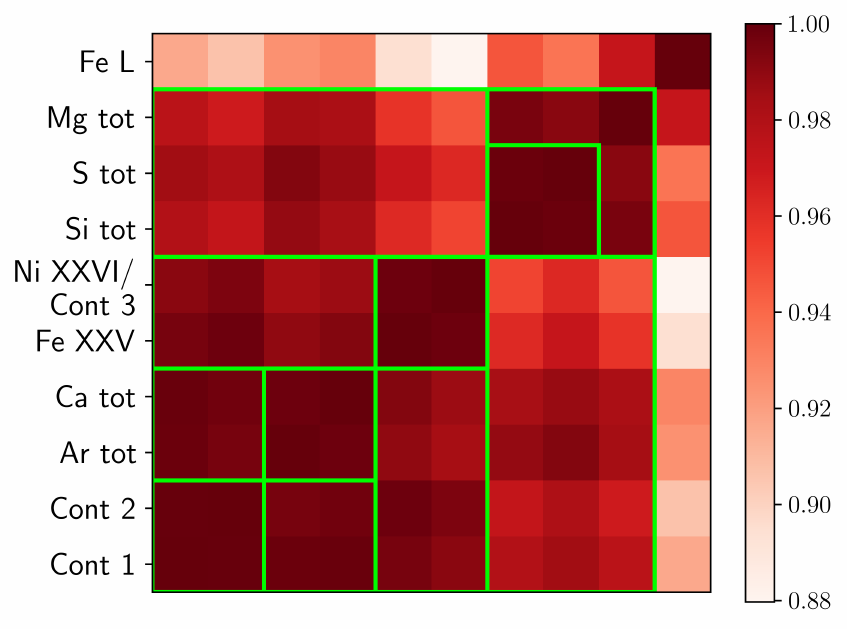}

\includegraphics[width=0.7\columnwidth]{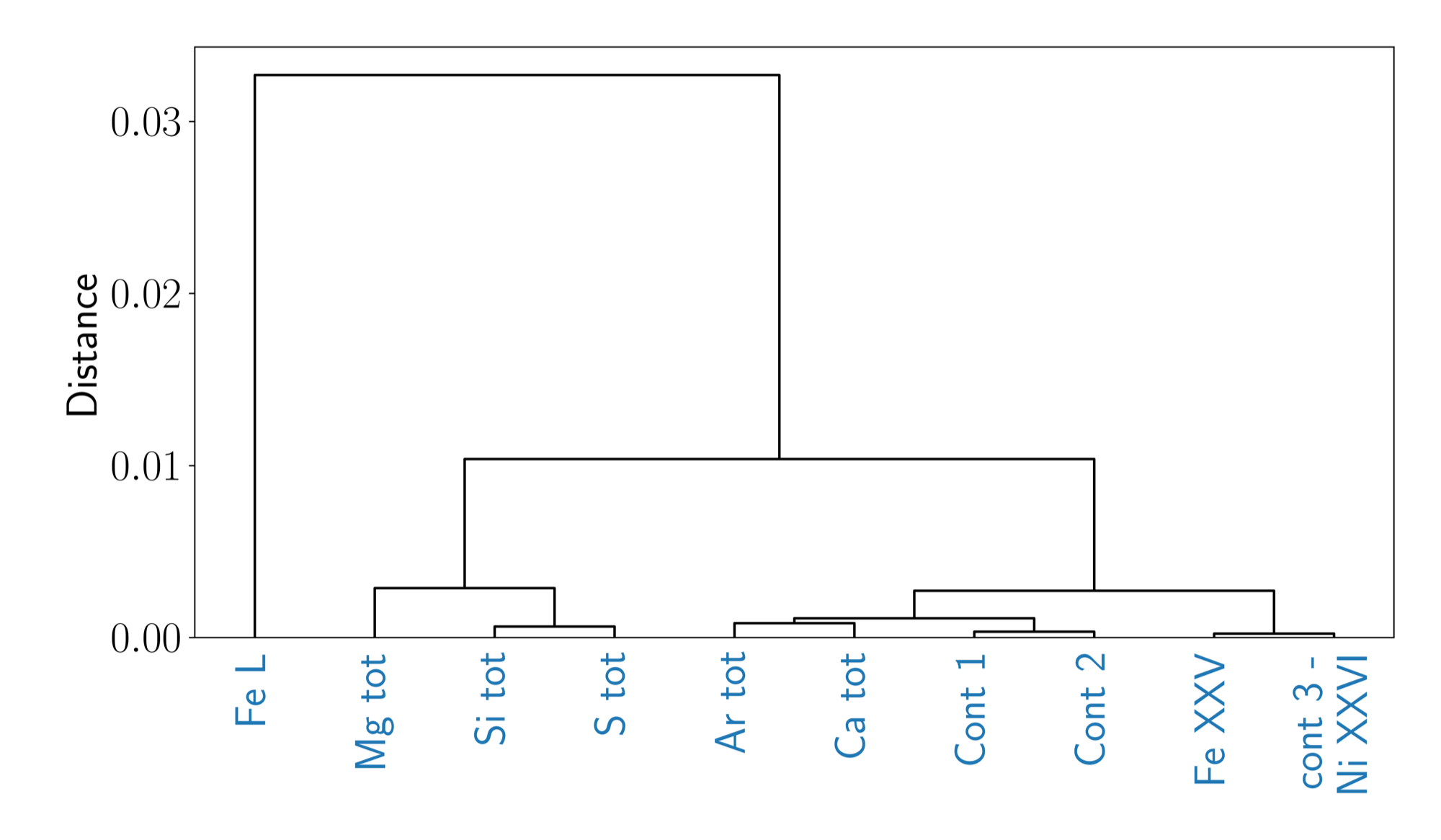}
    \caption{Analysis of the flux maps synthesized from the HD simulations of Cassiopeia A by \cite{2022A&A...666A...2O}. 
    The upper panel shows the dendrogram obtained by using the SVHC method with average linkage and the Frobenius distance of Eq. (\ref{frob}). 
    The central panel shows the distance matrix reordered according to the previous dendrogram.
    The lower panel shows the one obtained using the average linkage and the cosine distance of Eq. (\ref{cos}).}
    \label{fig:CASA-flux}
\end{figure}

A first investigation was done by analyzing the synthetic maps derived from the dedicated 3D HD simulations of Cassiopeia A by \cite{2022A&A...666A...2O} in order to assess whether our methodology is effective or not.
In Fig. \ref{fig:CASA-flux} we show the dendrograms obtained using the clustering methodologies illustrated in Section \ref{HT}. The upper panel shows the dendrogram obtained using the SVHC  method with the Frobenius distance (see Eq. \ref{frob}), while the lower panel shows the one obtained using the cosine distance (Eq. \ref{cos}). The central panel shows the distance matrix, which was reordered according to the hierarchical dendrogram obtained using the SVHC methodology. Results are nicely consistent with each other, thus indicating that they are quite robust against the nonlinearity introduced in the distance, $d_{ij}=\sqrt{2 \, (1 - \rho_{ij})}$. 

We point out that in the HD simulation Fe and Ni are synthesized in the very same ejecta layers. This is in agreement with the short distance between the Fe XXV and Ni XXVI maps shown in Fig. \ref {fig:CASA-flux}. Similar considerations can be drawn for Si and S, which are clustered together in the dendrograms, as well as for Ar and Ca. These results indicate that our dendrograms correctly recover the spatial distribution of the metal-rich plasma.

\begin{figure} [htb]
    \centering
    \includegraphics[width=0.7\columnwidth]{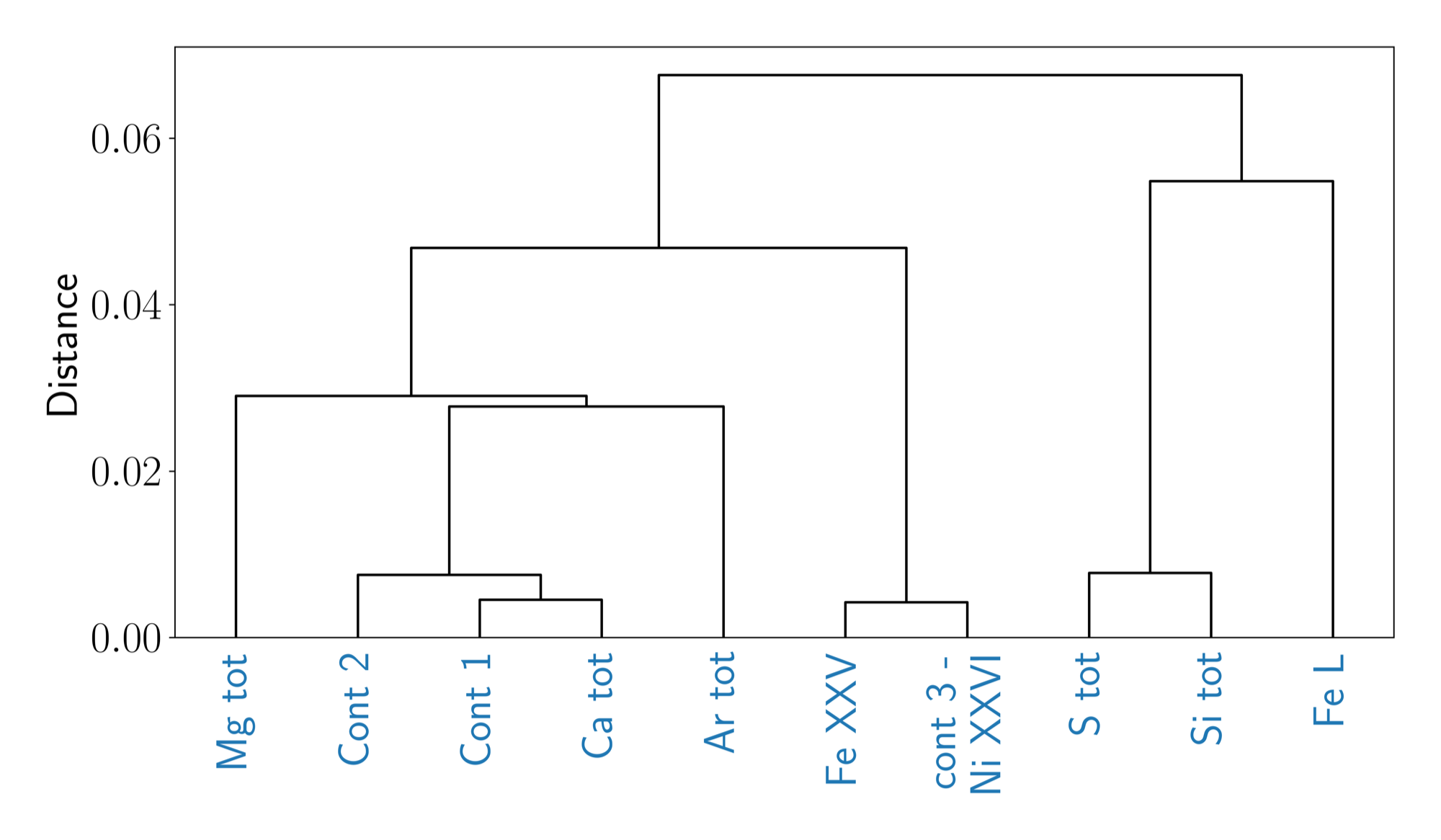}
    \caption{Analysis of the flux maps synthesized from the HD simulations of Cassiopeia A by \cite{2022A&A...666A...2O}. 
                  The panel shows the dendrogram obtained using the GNN embeddings ($k=6$), followed by the application of average linkage with Euclidean distance.} 
    \label{fig:casagnn3}
\end{figure}

The same results are observable when we consider the graph-based approach. In Fig.~\ref{fig:casagnn3} we report the graph we built and the results obtained using the graph-based approach, $k=6,$ and the cosine distance. Though similar results are obtained with other $k$ values (confirming that the clustering structures are preserved even after translating the original flux maps into a graph representation), $k=6$ provides the most striking agreement with the results shown in Fig. \ref{fig:CASA-flux} and was be adopted hereafter. The consistency between dendrogram-based techniques and the deep-learning graph-based approach indicates that the intrinsic spatial correlations characterizing the remnant are sufficiently strong to be captured not only through traditional HC techniques but also through a topological encoding of the data. In particular, the graph construction enhances local neighborhood relationships, allowing us to characterize the connectivity patterns among regions of similar flux intensities. The resulting graph-based clusters therefore reflect the same underlying physical organization observed in the map-based dendrograms, suggesting that the emission morphology is robust against changes in the representation space.

\begin{figure} [htb]
    \centering
    
    \includegraphics[width=0.7\columnwidth]{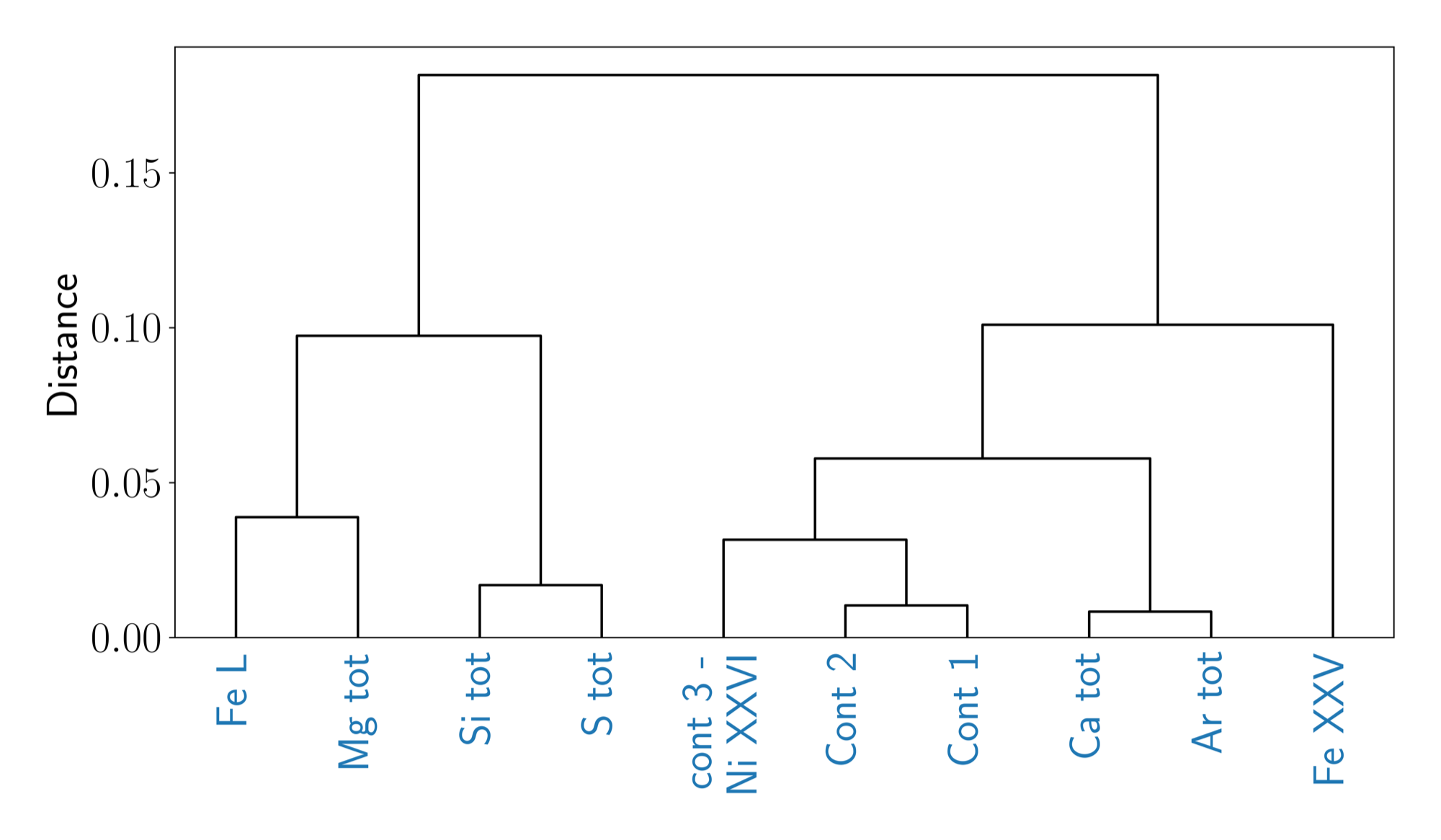}
    \caption{Analysis of the \emph{Chandra} X-ray flux maps of Cassiopeia A.
                  The panel shows the dendrogram obtained using the average linkage with the cosine distance of Eq. (\ref{cos}).}
    \label{fig:casareal}
\end{figure}

It is interesting to compare the results described above for the HD simulation of Cassiopeia A with those that we obtained by analyzing the actual \emph{Chandra} observations of Cassiopeia A. As an example, Fig. \ref{fig:casareal} shows the dendrogram of the Cassiopeia A \emph{Chandra} maps produced using the average linkage method with the cosine distance. The similarity with the dendrograms derived from the HD simulations (Fig. \ref{fig:CASA-flux} and Fig. \ref{fig:casagnn3}) is impressive, confirming the good agreement between model and observations and, most importantly, offering an important key to interpreting the observed features, as discussed in Sect.~\ref{concl}. 

\begin{figure}[htb] 
    \centering 
    \includegraphics[width=0.7\columnwidth]{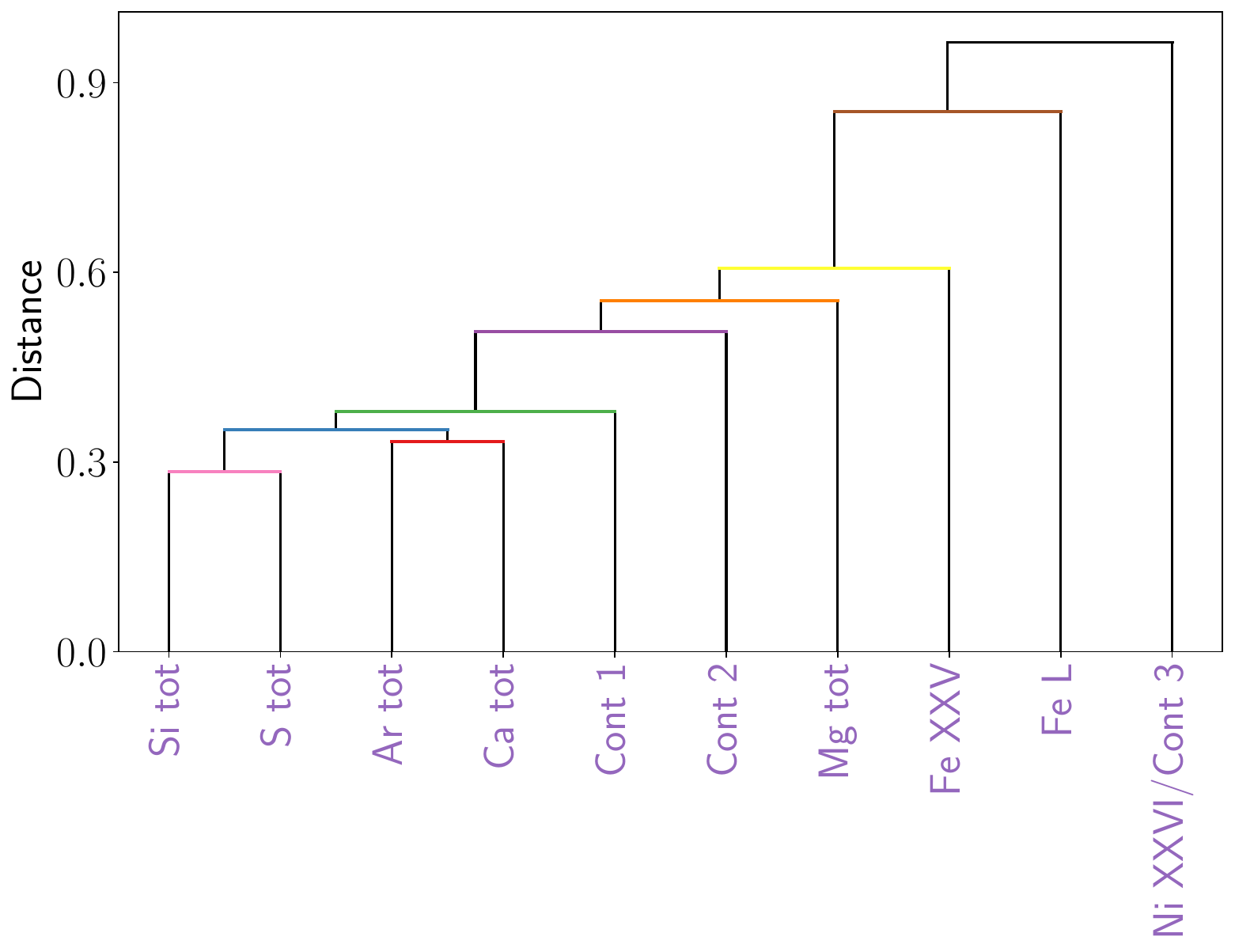} 
    \includegraphics[width=0.7\columnwidth]{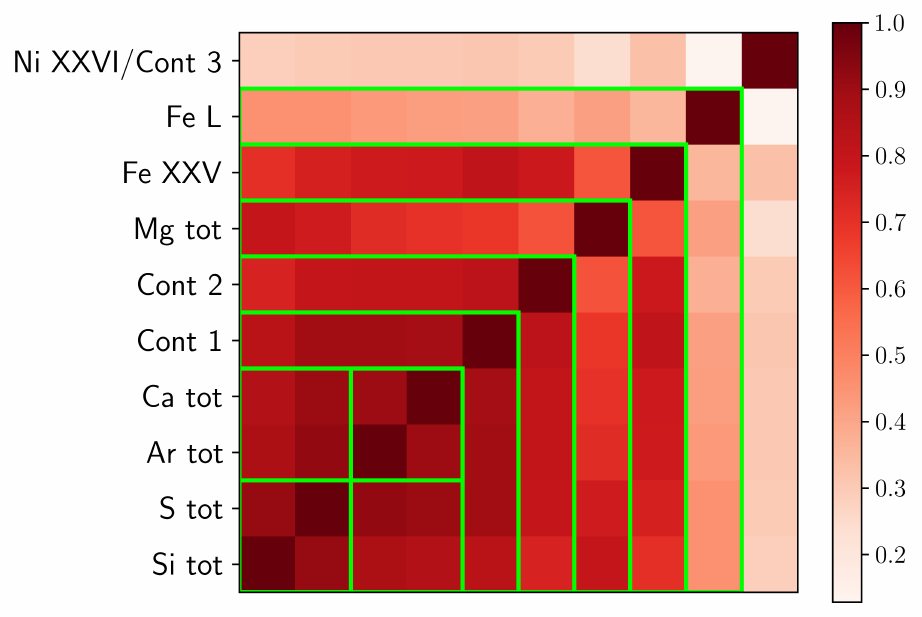}
    \includegraphics[width=0.7\columnwidth]{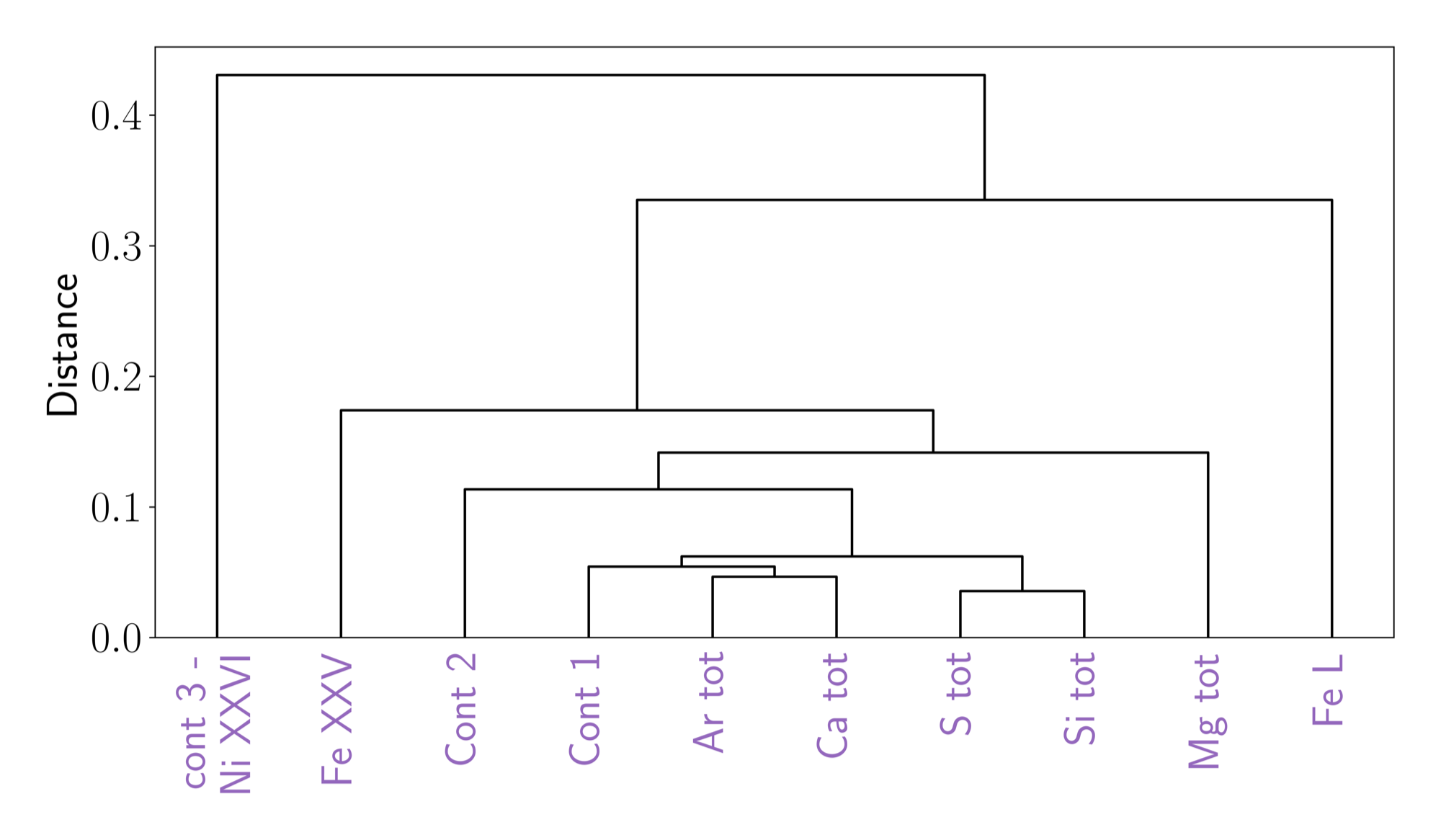}
    \caption{Analysis of the \emph{Chandra} X-ray flux maps of W49B.
    The upper panel shows the dendrogram obtained using the SVHC methods with average linkage and the Frobenius distance of Eq. (\ref{frob}). 
    The central panel shows the distance matrix reordered according to the previous dendrogram.
    The lower panel shows the one obtained using the average linkage and the cosine distance of Eq. (\ref{cos}).}
    \label{fig:W49B-flux-empirical}
\end{figure}

Overall, the above results for Cassiopeia A are confirmed when considering W49B. Similarly to what has already been done, in Fig. \ref{fig:W49B-flux-empirical} we show the dendrograms obtained using the two clustering methodologies illustrated in Section \ref{HT}. Also in this case, Si and S are clustered together, as are Ar and Ca. We also notice how the Fe XXV map stands separate from intermediate-mass elements, revealing a remarkably different distribution of the Fe-rich ejecta. This confirms that our methodology can efficiently identify the common morphogenetic origin of the elements in SNRs.

\begin{figure}[htb]
    \centering
    \includegraphics[width=0.7\columnwidth]{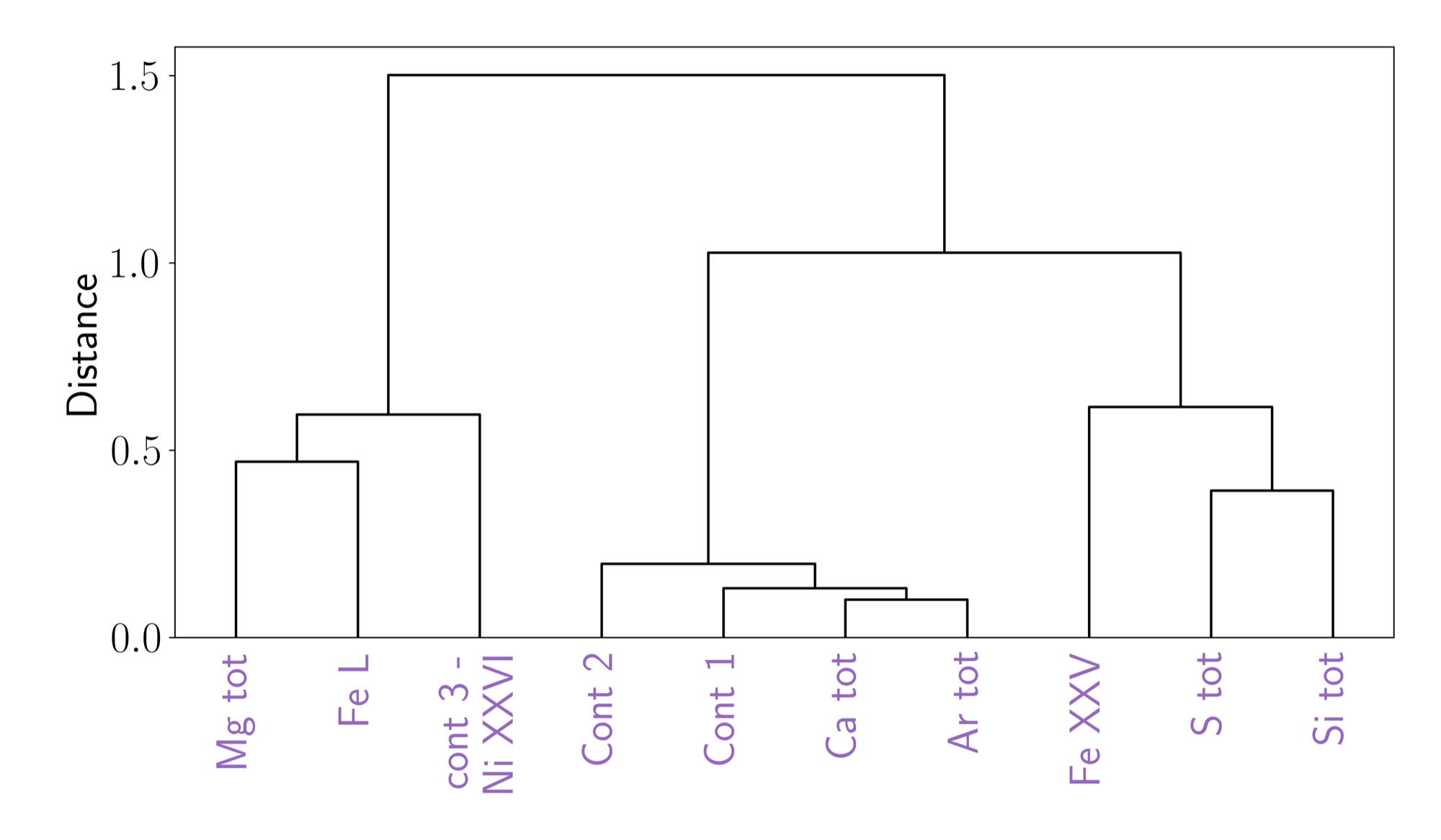}
    \caption{Analysis of the \emph{Chandra} X-ray flux maps of W49B.
                  The panel shows the dendrogram obtained using the GNN embeddings ($k=6$), followed by the application of average linkage with Euclidean distance.}
    \label{fig:w49gnn}
\end{figure}

Moreover, the graph-based analysis (see Fig.~\ref{fig:w49gnn}), performed using $k=6$ and the Euclidean distance, reproduces the same hierarchical structures observed in the dendrograms. The agreement between the dendrogram-based and graph-based approaches further demonstrates that our framework provides a reliable and stable tool for identifying coherent structures in flux maps, independent of the specific representation or similarity metric used.
The dendrograms produced for the other remnants with the two cluster methodologies and the graph-based analysis are all shown in Appendix~\ref{app:A}.

\subsection{Multiple SNR investigation} \label{resultsfluxmulti}

\begin{figure*}[htb] 
         \sidecaption
 \begin{minipage}{12cm}  
    \includegraphics[width=\textwidth]{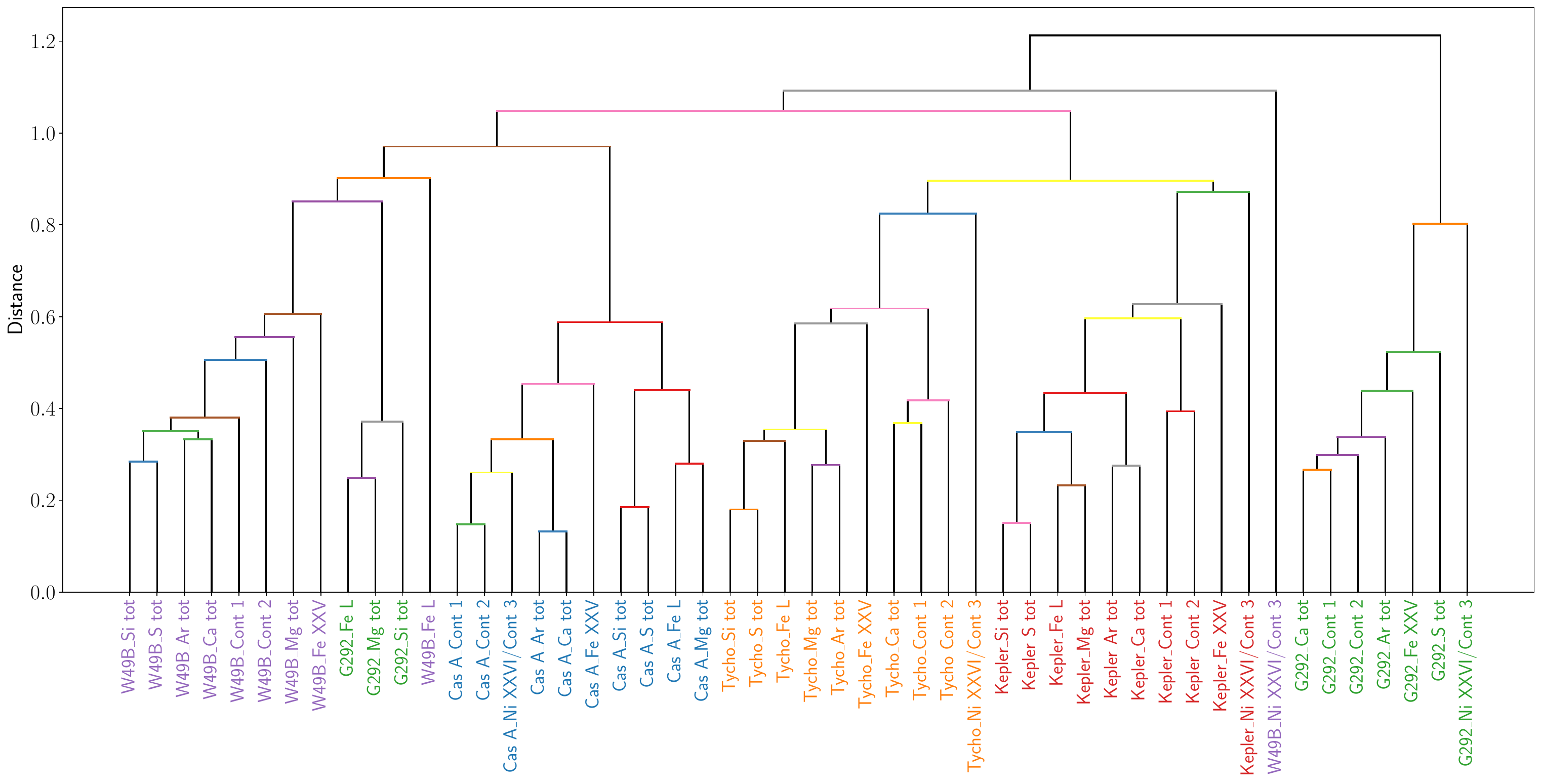}
    \includegraphics[width=\textwidth]{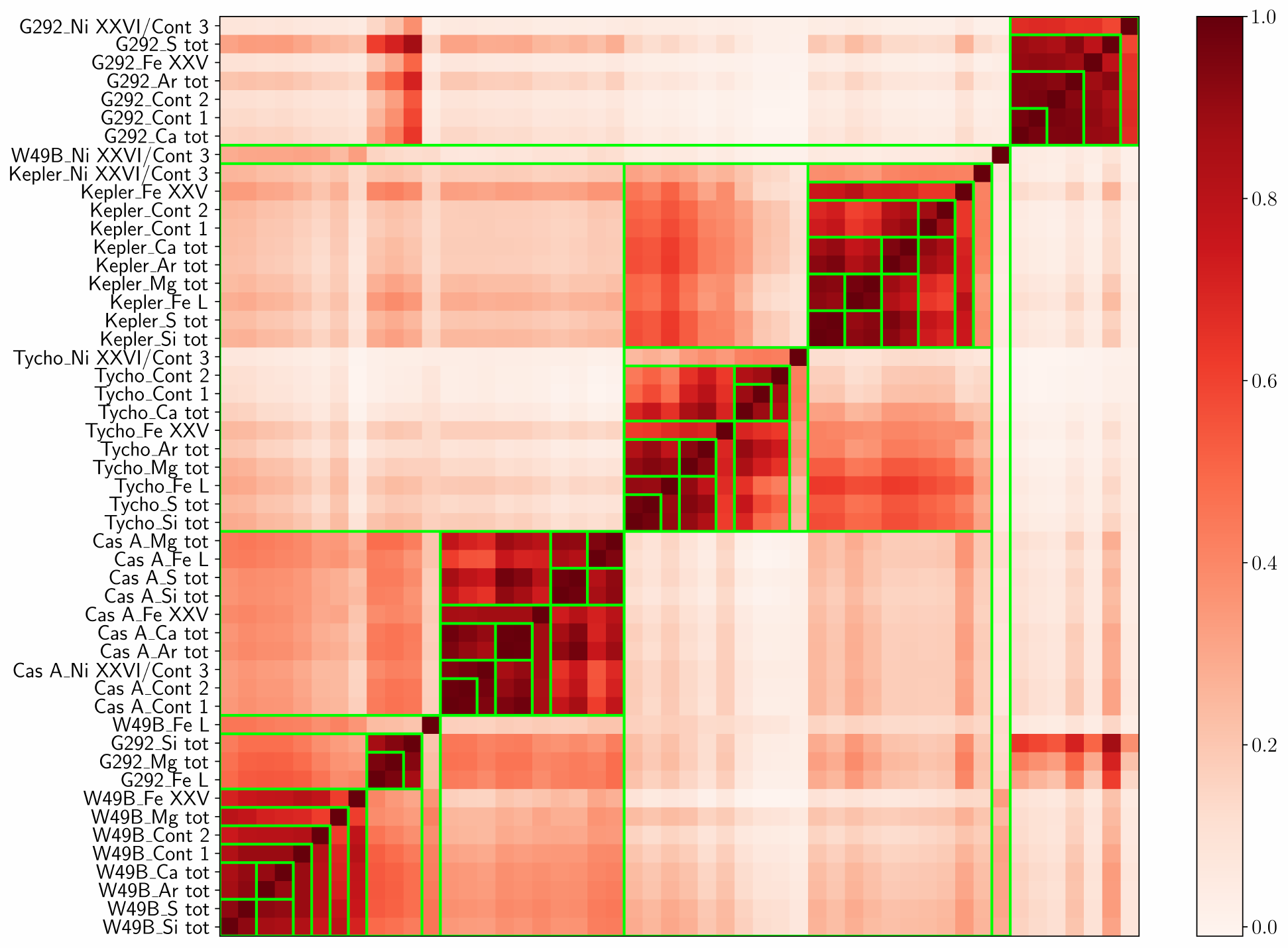}
   \includegraphics[width=\textwidth]{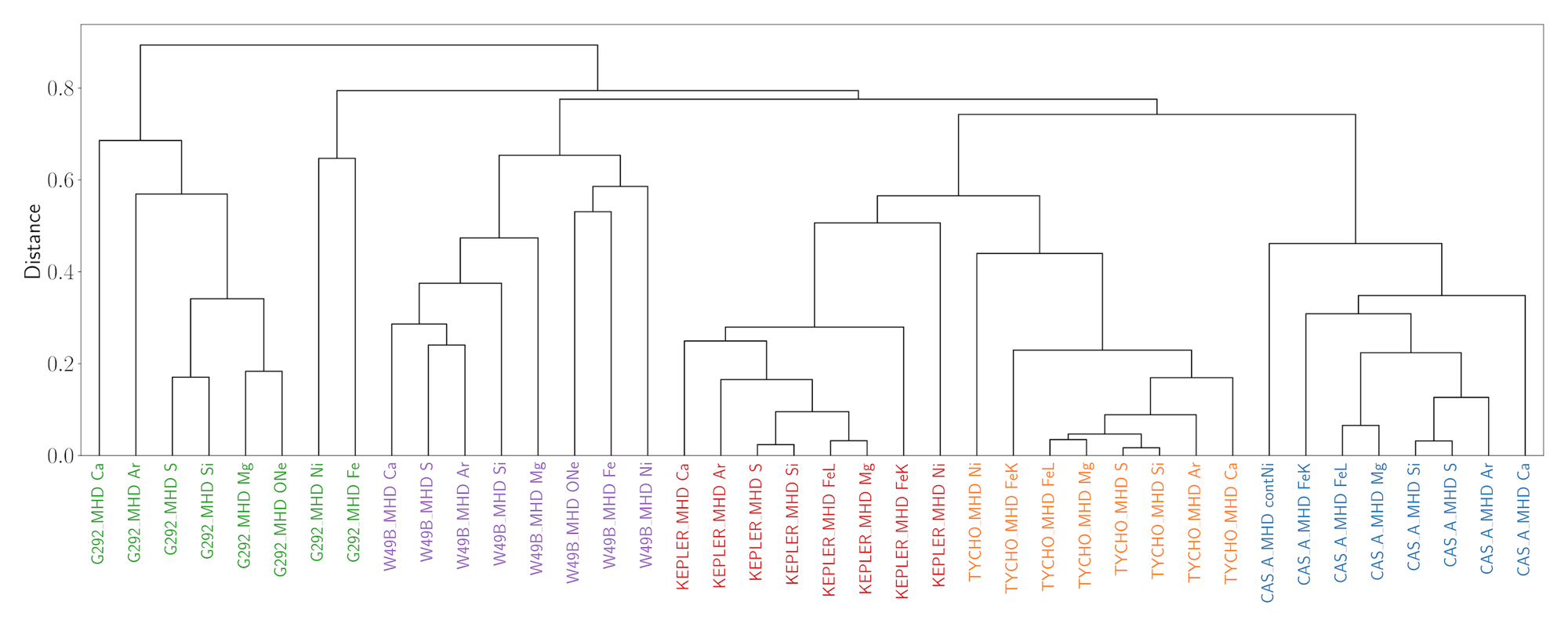}
    \end{minipage}
    \caption{Analysis of the \emph{Chandra} X-ray flux maps of all the remnants analyzed in this paper. The upper panel shows the dendrogram obtained using the SVHC methods with average linkage and the Frobenius distance of Eq. (\ref{frob}). The central panel shows the distance matrix reordered according to the previous dendrogram. The lower panel shows the one obtained by using the average linkage and the cosine distance of Eq. (\ref{cos}).} 
    \label{fig:all-flux}
\end{figure*}

We used dendrograms to try to understand whether we can infer similarities amongst X-ray maps of different SNRs. We constructed the dendrograms starting from the two $50 \times 50$ similarity matrices obtained using the Frobenius distance of Eq. (\ref{frob}) and the cosine distance of Eq. ($\ref{cos}$). Such dendrograms are shown in the upper and lower panels of Fig. \ref{fig:all-flux}, respectively. Both dendrograms show that the main driver of the clustering is the membership to a given SNR. However, the hierarchies shown in the dendrograms are still quite informative. In particular, it is worth noticing that the X-ray narrow-band images of Tycho's SNR and Kepler's SNR are quite close to each other. The SVHC dendrogram shows that the link between the two groups of elements is colored in yellow, which would indicate that this connection is statistically significant, and therefore Tycho's SNR and Kepler's SNR form a statistically significant cluster. 

On the other hand, most of the W49B and three G292.0+01.8 elements form a statistically significant cluster that is then linked in a statistically significant way to most of the Cassiopeia A maps. The remaining G292.0+01.8 elements form an extra statistically significant cluster (the rightmost cluster in the SVHC dendrogram and the orange cluster in the cosine distance dendrogram). Such a cluster, however, is not linked in a statistically significant way with the remaining part of the dendrogram.  

\begin{figure*}[htb] 
    \sidecaption 
    \includegraphics[width=0.65\textwidth]{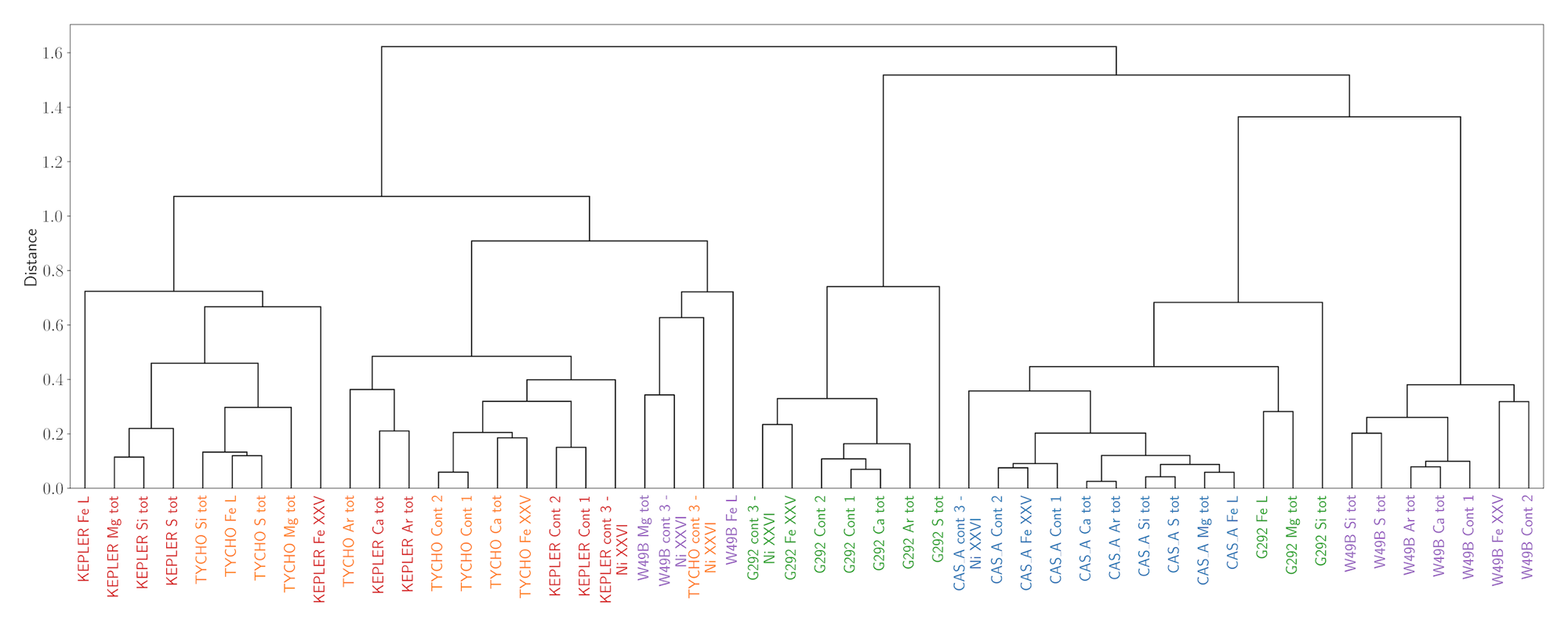}
    \caption{Analysis of the \emph{Chandra} X-ray flux maps of all the remnants analyzed in this paper.
      The panel shows the dendrogram obtained using the GNN embeddings ($k=6$), followed by the application of average linkage with Euclidean distance.}
    \label{fig:GNNdendro}
\end{figure*}

In Fig.~\ref{fig:GNNdendro} we report the results obtained with the graph-based approach. The dendrogram obtained is comparable to those shown in Fig.~\ref{fig:all-flux}
both for the clustering of flux maps within a given SNR and among different remnants. 
In particular, we find that W49B is more similar to Cassiopeia A  and G292.0+01.8, while Tycho's SNR and Kepler's SNR are more similar to each other. 

\begin{figure*}[htb] 
    \sidecaption
    \begin{minipage}{12cm}
    \includegraphics[width=\textwidth]{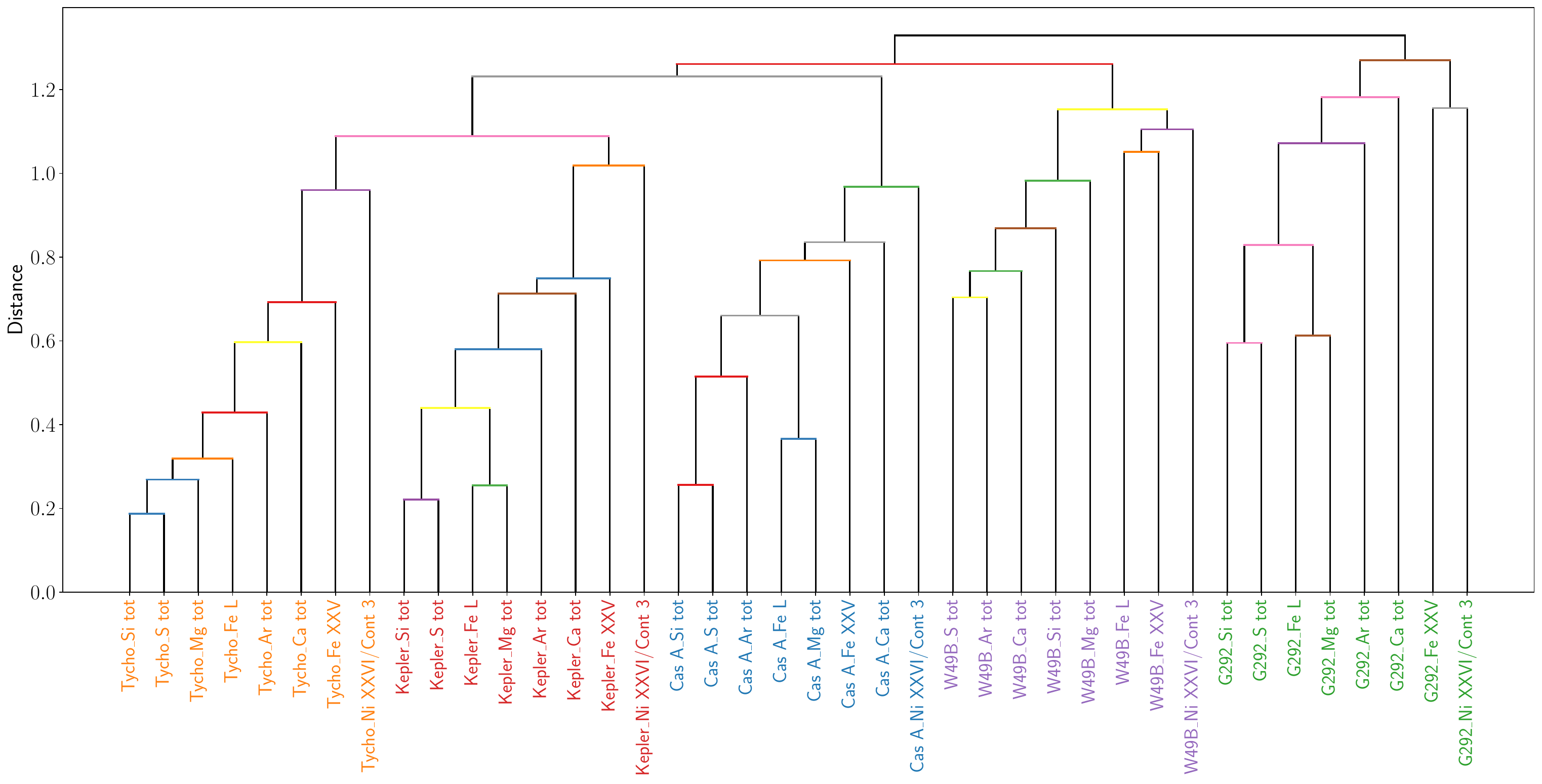}
    \includegraphics[width=\textwidth]{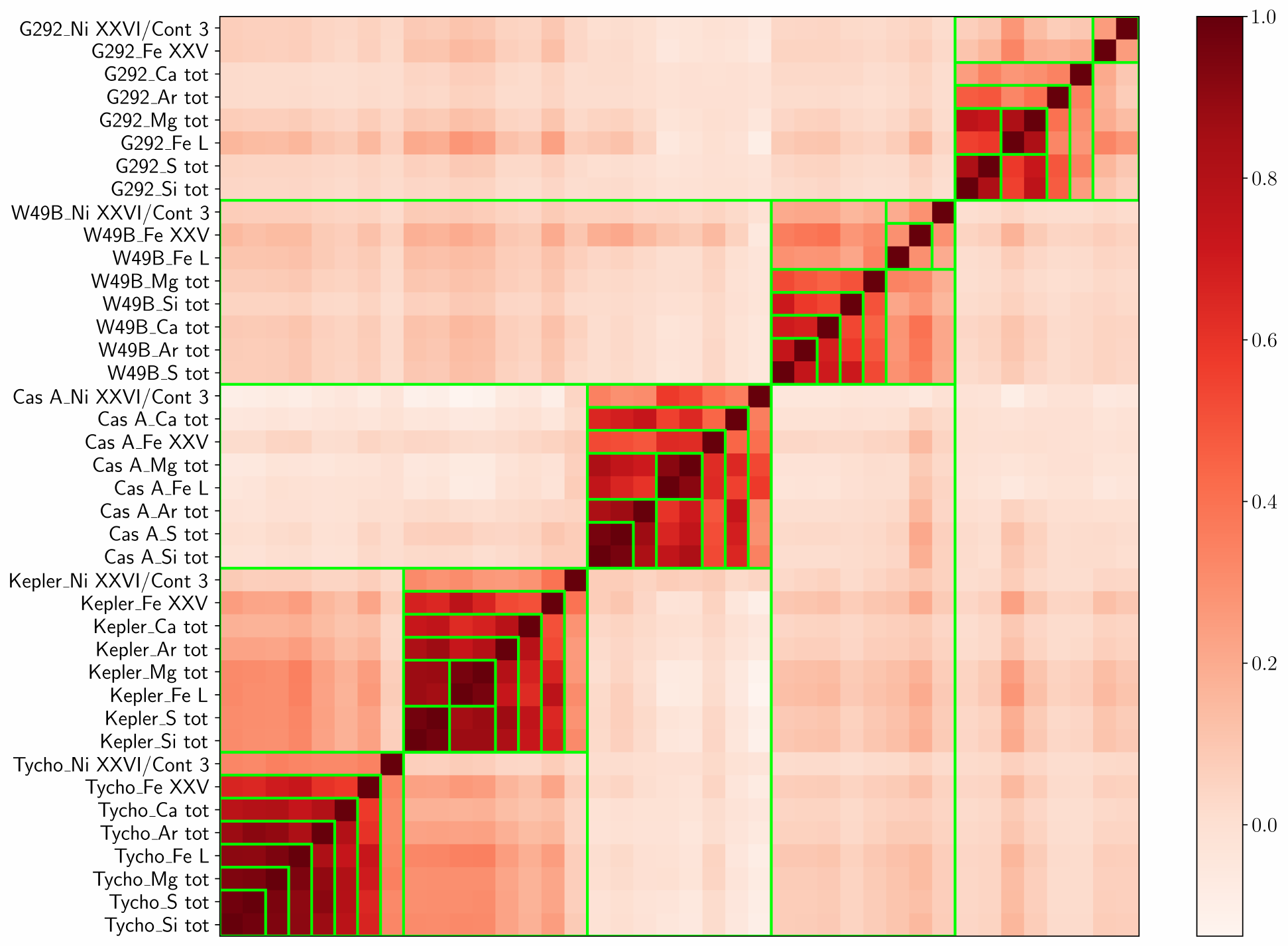}
    \includegraphics[width=\textwidth]{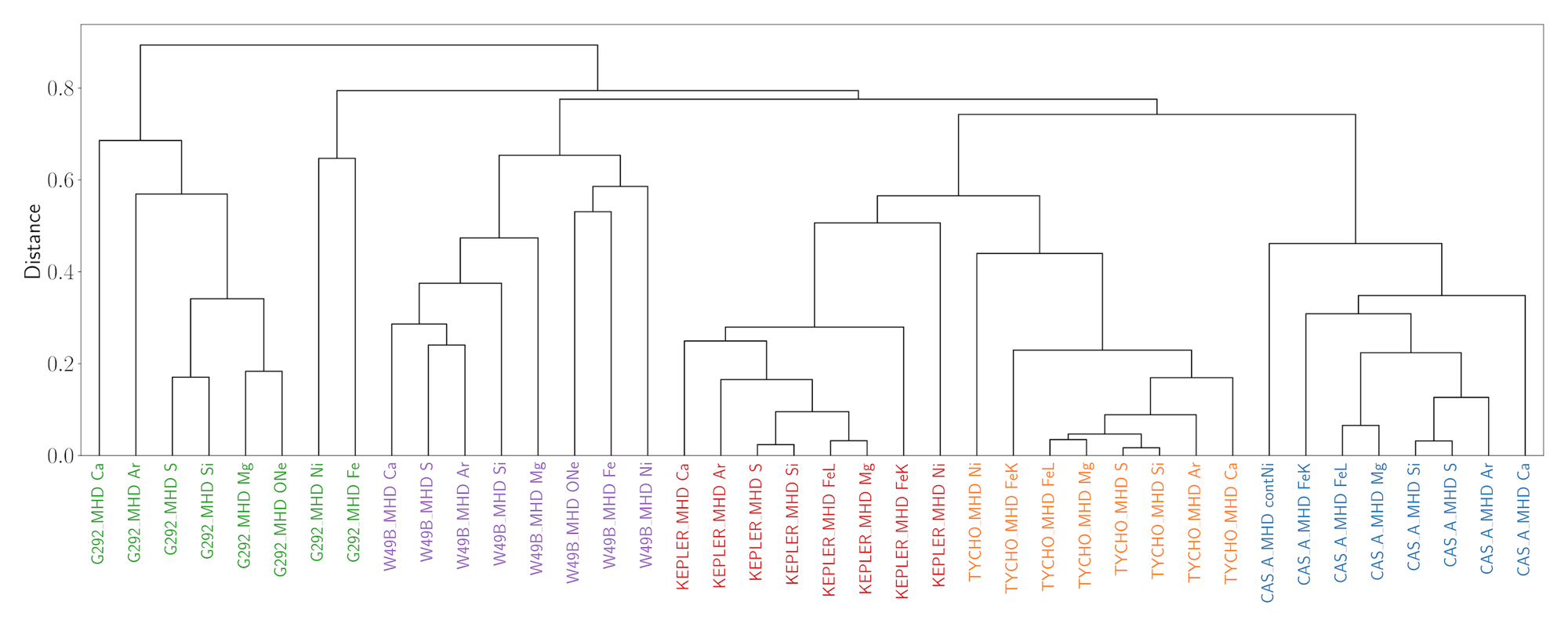}
        \end{minipage}
    \caption{Analysis of the \emph{Chandra} X-ray EW maps of all the remnants analyzed in this paper. 
    The upper panel shows the dendrogram obtained using the SVHC methods with average linkage and the Frobenius distance of Eq. (\ref{frob}). 
    The central panel shows the distance matrix reordered according to the previous dendrogram.
    The lower panel shows the one obtained using the average linkage and the cosine distance of Eq. (\ref{cos}).}    \label{fig:all-ew}
\end{figure*}

The above results are further supported by the analysis of the EW maps Though it is important to note that EW maps also depend on temperature, absorption column density, and ionization timescale, they can be informative tracers of the spatial distribution of the chemical properties of the plasma. In this case, the continuum band (i.e., 4.0–5.0 keV) was obviously not taken into account. 
We constructed the dendrograms starting from the two $40 \times 40$ similarity matrices obtained using the Frobenius distance of Eq. (\ref{frob}) and the cosine distance of Eq. ($\ref{cos}$).
Such dendrograms are shown in the upper and lower panels of Fig. \ref{fig:all-ew}, respectively. 
Analogously to the case of flux maps, dendrograms show that the main driver of the clustering is still the membership to a given SNR. Also in this case, the hierarchical structure of the dendrograms is informative, revealing the significant similarity between Tycho's SNR and Kepler's SNR, while most of the W49B maps are closer to those of G292.0+01.8 and Cassiopeia A, rather than to those of Type Ia SNRs.

When considering the GNN-based approach, the results are consistent with those obtained from the classical HC analyses, as shown in Fig.~\ref{fig:all-ewGNN}, in which we report the related dendrogram obtained using $k=6$ for the graph construction. 

\subsection{Computational costs}
From a computational perspective, the cosine-distance method requires computing a pairwise distance matrix, followed by HC. SVHC is more computationally expensive due to its bootstrap-based validation procedure, which provides statistically supported clusters and corresponding p values. When comparing the cosine-based distance approach and the GNN-based approaches, the GNN-based one represents the highest computational cost, as it involves graph construction and neural-network training to learn refined node representations. However, since training is performed on a single graph, the method can benefit substantially from parallel processing and GPU acceleration, reducing the practical impact of its computational complexity.

For this purpose, we report the computation times in milliseconds for the three approaches in our analysis. As a benchmark for computational-cost evaluation, we selected the experiment involving 50 spectral bands, as it represents the largest dataset considered in this study and therefore constitutes the most demanding scenario in terms of processing requirements. For fairness, all the experiments have been performed on the same workstation equipped with an Intel Core i7-8700 3.2 GHz CPU and an Nvidia Titan V GPU. The total system memory is 32 GB of DDR4, and the GPU supplies 12 GB of DDR5 memory. The operating system is Ubuntu 24.4.04. For the cosine-based distance approach, the elapsed time is 16962.7 ms; i.e., 1365064.8 ms  for the SVHC methodology and 15937.0 ms for the GNN-based approach.

\begin{figure*}[htb] 
    \sidecaption
    \includegraphics[width=12cm]{DENDROGRAMS/GNN_EW6.png}
    \caption{Analysis of the \emph{Chandra} X-ray EW maps of all the remnants analyzed in this paper. 
                  The panel shows the dendrogram obtained using the GNN embeddings ($k=6$), followed by the application of average linkage with Euclidean distance.}    \label{fig:all-ewGNN}
\end{figure*}

\section{Discussion and conclusions} 
\label{concl}

We developed two dendrogram-based techniques (a multivariate analysis and a deep-learning graph-based approach) to quantitatively investigate similarities and differences among X-ray maps of SNRs.
The two techniques provide essentially the same results. We consider this a highly positive outcome, because the two methodologies counter-check each other's results, thus increasing the robustness of our findings. In summary, the reason for presenting both approaches is twofold. On one side, GNN provides a complementary validation of the simpler clustering approach. On the other side, the validity of the training procedures required by the GNN approach can be assessed by a comparison with the simpler, classical clustering approach.

The results presented in Sect.~\ref{results} provide important clues for the analysis of these maps.
We first focus on the similarity between the dendrograms obtained for the narrow-band images synthesized from the numerical simulation of Cassiopeia A (Fig.~\ref{fig:CASA-flux} and Fig. ~\ref{fig:casagnn3}) and those obtained for the actual \emph{Chandra} observation of the same remnant (Fig. ~\ref{fig:casareal}). We can disentangle the role of abundances, temperatures, and ionization parameters in determining the similarities/differences among different maps. The similarity between Si and S, as well as that between Ar and Ca, shown in Fig. \ref{fig:casareal}, is remarkably congruent with that observed in the HD simulations, where it is the result of the spatial co-location of these elements. We can then associate the small distances between Si-S and Ar-Ca, with the spatial distribution of the ejecta chemical composition. In the same framework, we stress that the Fe XXV emission, which somehow carries memories of the pristine anisotropies in the inner layers of the ejecta, stands separate from maps of intermediate-mass elements (in the actual data, the Ni XXVI line is so weak that it is difficult to obtain information on the Ni distribution in its map). We finally noticed the similarity between the Fe L and Mg maps, which we interpret as an effect of the temperature distribution, reflecting the fact that these emission lines stem from plasma at lower temperatures with respect to other lines.

Similar considerations can also be drawn for W49B (Fig. ~\ref{fig:W49B-flux-empirical} and Fig.~\ref{fig:w49gnn}). In particular, we find that the spatial distribution of Fe-rich ejecta is significantly different from that of lighter elements.

The comparative analysis of X-ray maps extracted from different remnants shows that our methodology is also capable of spotting analogies (differences) that might be related to the common (different) physical origin of the object under examination. 
In particular, in Sect.~\ref{resultsfluxmulti} we provide proof that Tycho's SNR and Kepler's SNR form a statistically significant cluster. This result might be ascribed to the fact that Tycho's SNR and Kepler's SNR share a common origin with Type Ia SNRs. 
Moreover, both the analysis of narrow-band images and EW maps reveal the similarity between the two core-collapse SNRs, namely Cassiopeia A and G292.0+01.8. It is worth noticing that it is not easy to spot this similarity by eye, while our methodology objectively identifies and separates the cluster of core-collapse SNRs from that of Type Ia SNRs.

We also notice that, albeit only in some cases, all the maps of G292.0+01.8 are not clustered together but distributed over two different branches (see Fig. \ref{fig:all-flux} and Fig. \ref{fig:GNNdendro}). This might be related to the fact that, while all the other remnants are ejecta-dominated, in G292.0+01.8 the X-ray emission stems both from the ejecta and the shocked ambient medium, and this might cause some differences among the line images (and the corresponding EW maps). Specifically, we note that Si, Mg, and Fe L form a branch that is separated from that including S, Ar, Ca, the Fe XXV line emission and, notably, the continuum. This may suggest a different origin for these two groups. Mg-rich, Si-rich and Fe-rich knots have been observed at different positions in G292.0+01.8, while, to our knowledge, ejecta knots characterized by enhanced Ar, Ca and Fe XXV emission have not been observed (e.g., \citealt{pbs19}). This may support an association of the branch including Si, Mg, and Fe L with the ejecta, and of the other one with the shocked ambient medium. However, we caution that further analysis is necessary to prove this scenario. In particular, including in our dendrogram maps of the O (and Ne) line emission can be highly informative, given that  G292.0+01.8 is O-rich, but such analysis goes beyond the scope of this paper. In any case, we notice that even if G292.0+01.8 is not ejecta-dominated its similarity with Cassiopeia A is still recovered by our dendrograms, showing our capability to spot objects with the same physical origin.

Interestingly, we found out that W49B, whose origin is still debated, is closer to core-collapse SNRs than to Type Ia SNRs. Though morphological similarity alone cannot settle the progenitor question,
these results might suggest that the morphogenesis of W49B is closer to that of G292.0+01.8 and Cassiopeia A rather than to Tycho's SNR and Kepler's SNR.

We caution that, given the small set of remnants considered, our tool cannot yet be regarded as a robust SNR classifier, nor can we draw definitive conclusions regarding the origin of W49B. However, while currently validated on a limited sample size of a few remnants, the application of this tool for distinguishing between Type Ia and core-collapse SNRs suggests a potential for effective classification, yielding encouraging initial results that warrant further investigation. Future perspectives will include extending our methodology to a broader set of SNRs by exploiting the Chandra and XMM-Newton archives.

More generally, the results discussed above show that our tool can be used to address different questions related to the analysis of both synthetic and actual X-ray maps of SNRs. In particular, our methodology provides useful information to i) assess the role of abundances and of temperature and ionization parameters in shaping the observed morphology of narrow-band images and EW maps; ii) check whether an MHD/HD simulation of a SNR reproduces the observed hierarchy of elemental structures; iii) highlight the role of a shocked ISM/CSM in a given map of a SNR through its anomalous clustering with respect to ejecta-dominated maps; iv) identify peculiar remnants (e.g., W49B) as outliers relative to known SNR classes. Additionally, a possible application of our tool for classifying other types of SNR (e.g., remnants interacting/non-interacting with molecular clouds) or identifying similarities among other classes of extended X-ray sources (e.g., galaxy clusters) can be envisaged.

\begin{acknowledgements}

We thank the anonymous referee for their valuable and helpful comments and suggestions. This work is supported by the  Fondazione  ICSC, Spoke 3 Astrophysics and Cosmos Observations. National Recovery and Resilience Plan (Piano Nazionale di Ripresa e Resilienza, PNRR) Project ID CN\_00000013 "Italian Research Center on High-Performance Computing, Big Data and Quantum Computing" funded by MUR Missione 4 Componente 2 Investimento 1.4: Potenziamento strutture di ricerca e creazione di "campioni nazionali di R\&S (M4C2-19)" - Next Generation EU (NGEU) - Research Project: {\em{TEcniche non Convenzionali per l’anaLisi di immagini Astrofisiche}} (TECLA) {\tt{tecla.unipa.it}}. The software used to perform the above analyses is freely available at:
{\tt{https://data.tecla.unipa.it/SNRsClust/}}

\end{acknowledgements}

\bibliographystyle{aa}
\bibliography{biblio}

\begin{appendix}

\section{Other Supernova Remnants} 
\label{app:A}

In this appendix, we present the results obtained for the additional SNRs considered in our analysis: G292.0+1.8, Kepler's SNR, and Tycho's SNR. We report findings for both the flux and EW maps. . 

\begin{figure}[h!] 
    \centering 
    \includegraphics[width=0.7\columnwidth]{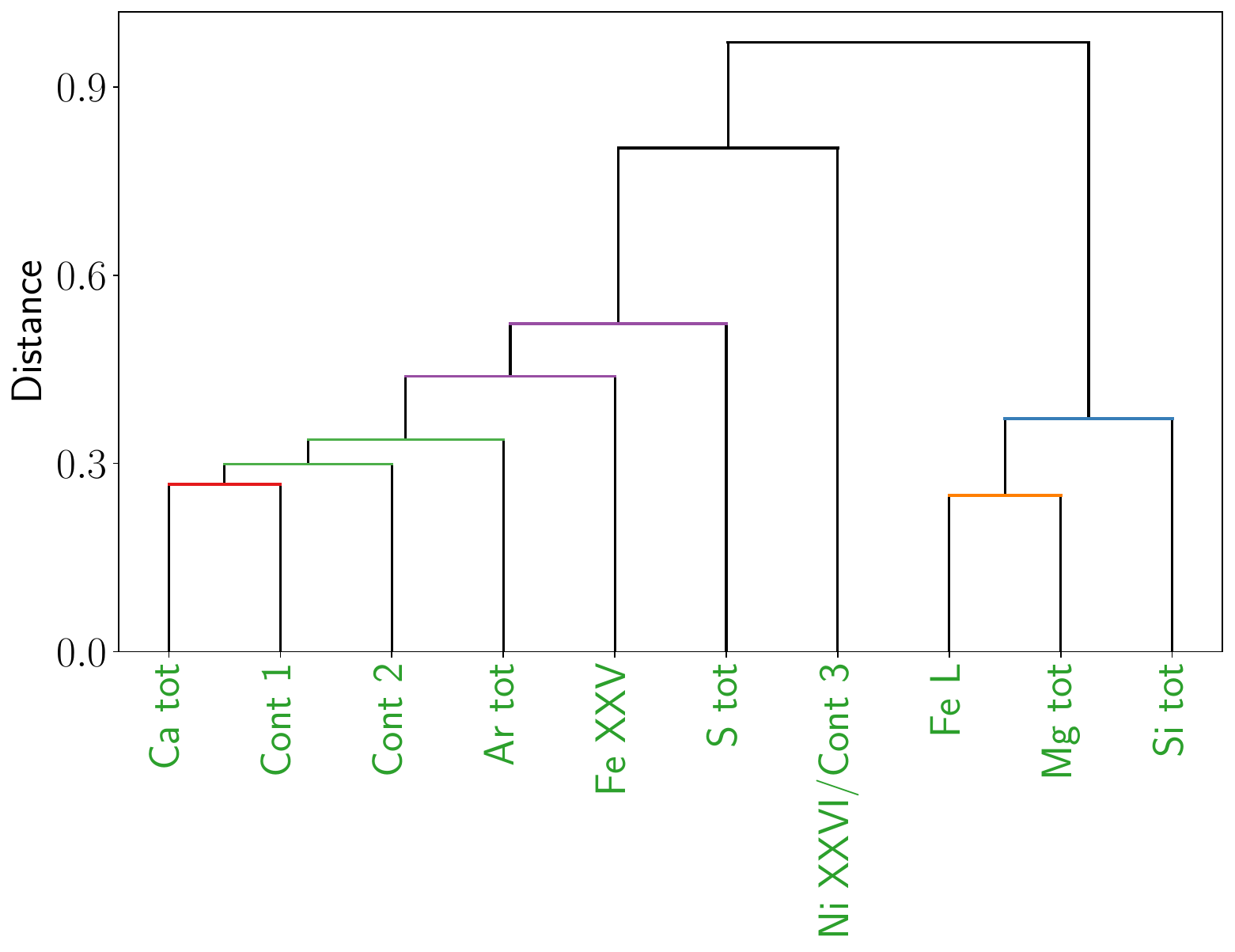}
    \includegraphics[width=0.7\columnwidth]{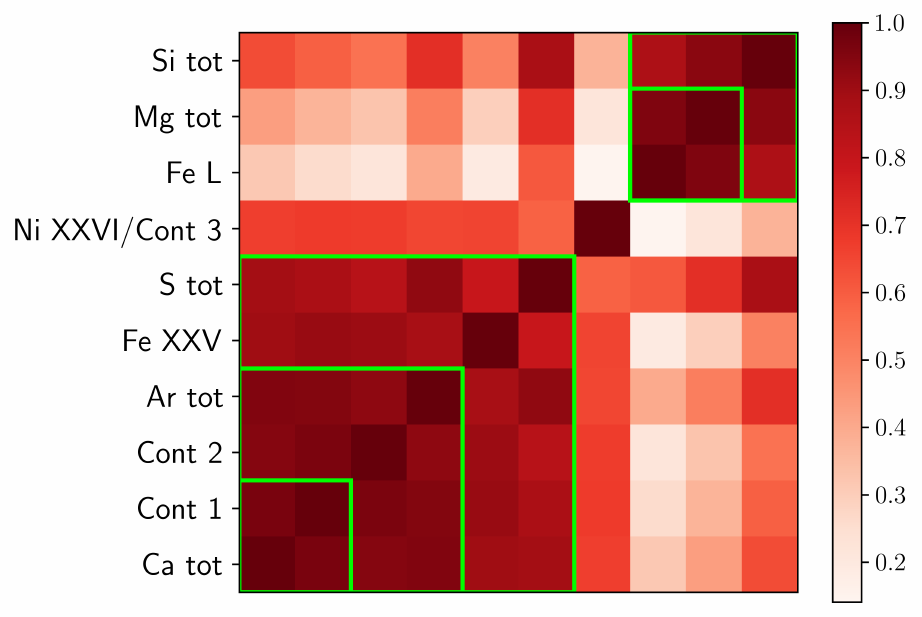}
    \includegraphics[width=0.7\columnwidth]{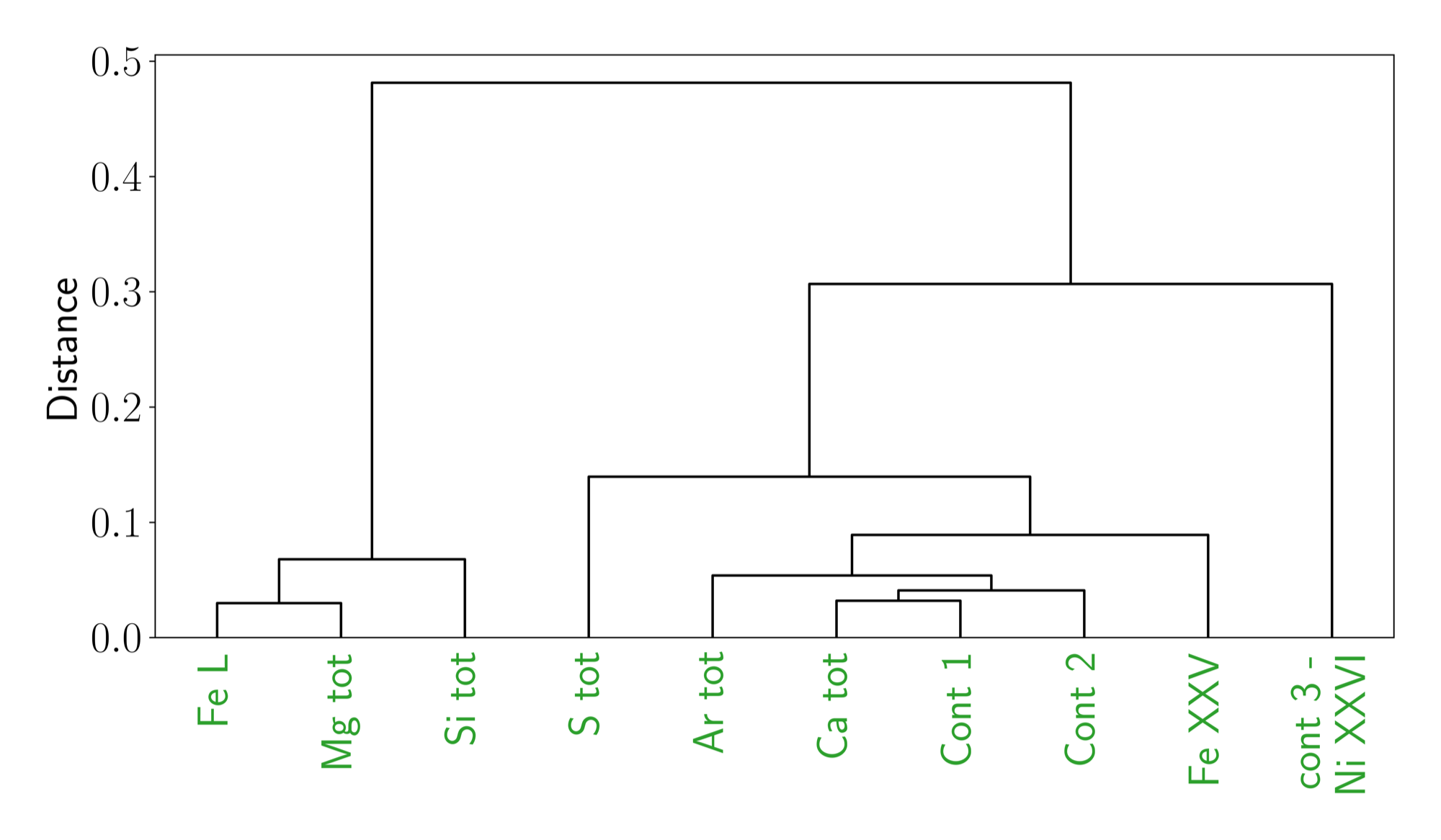}
    \caption{Analysis of the \emph{Chandra} X-ray flux maps of G292.
    The upper panel shows the dendrogram obtained by using the SVHC methods with average linkage and the Frobenius distance of Eq. (\ref{frob}). 
    The central panel shows the distance matrix reordered according to the previous dendrogram.
    The lower panel shows the one obtained by using the average linkage and the cosine distance of Eq. (\ref{cos}).}
    \label{fig:g292flux}
\end{figure}

\begin{figure}[htb] 
    \centering 
    \includegraphics[width=0.7\columnwidth]{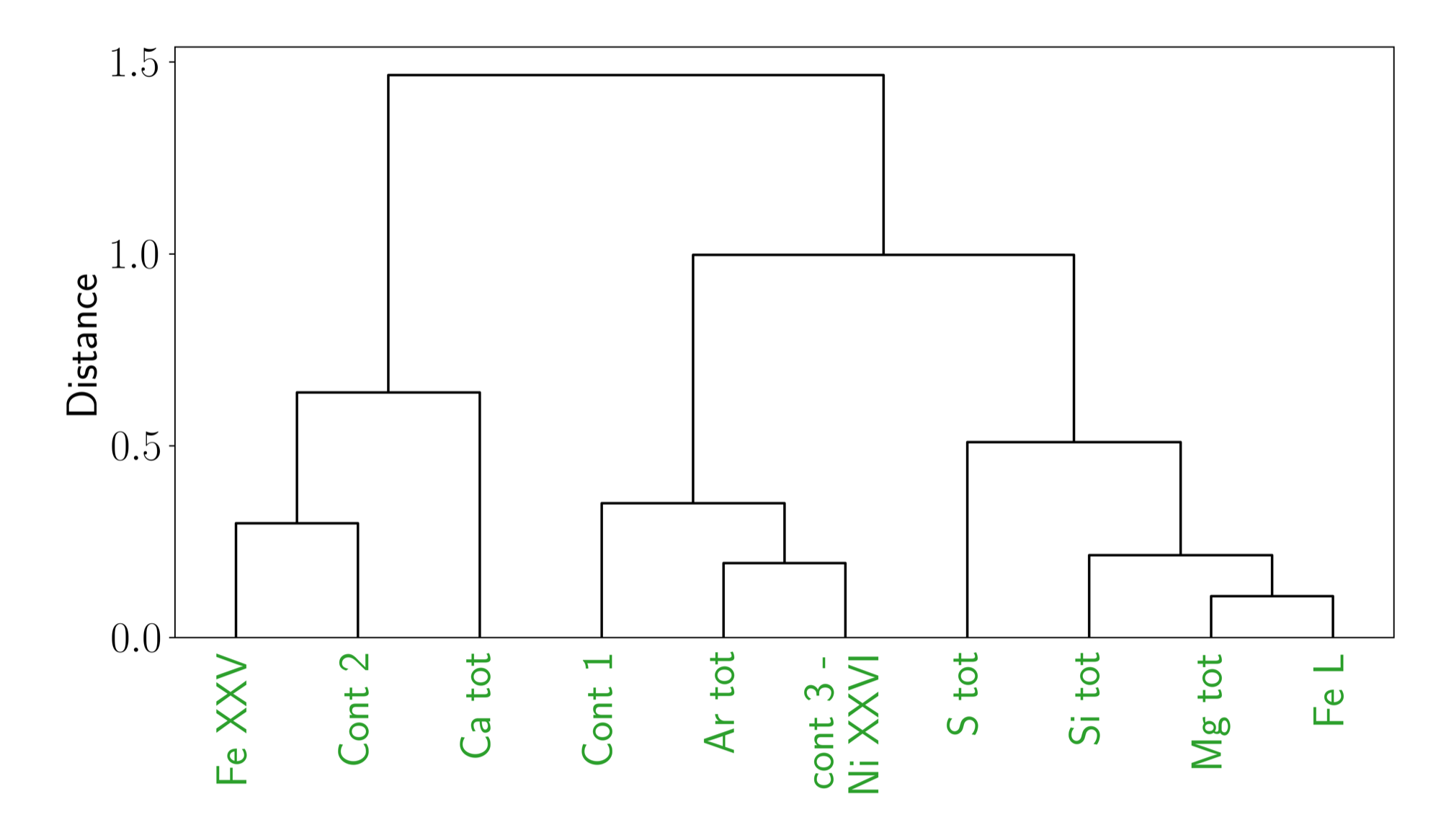}
    \caption{Analysis of the \emph{Chandra} X-ray flux maps of G292.0+1.8.
             The panel shows the dendrogram obtained by using the GNN embeddings ($k=6$), followed by the application of average linkage with Euclidean distance.}
    \label{fig:g292fluxGNN}
\end{figure}

\begin{figure}[htb] 
    \centering 
    \includegraphics[width=0.7\columnwidth]{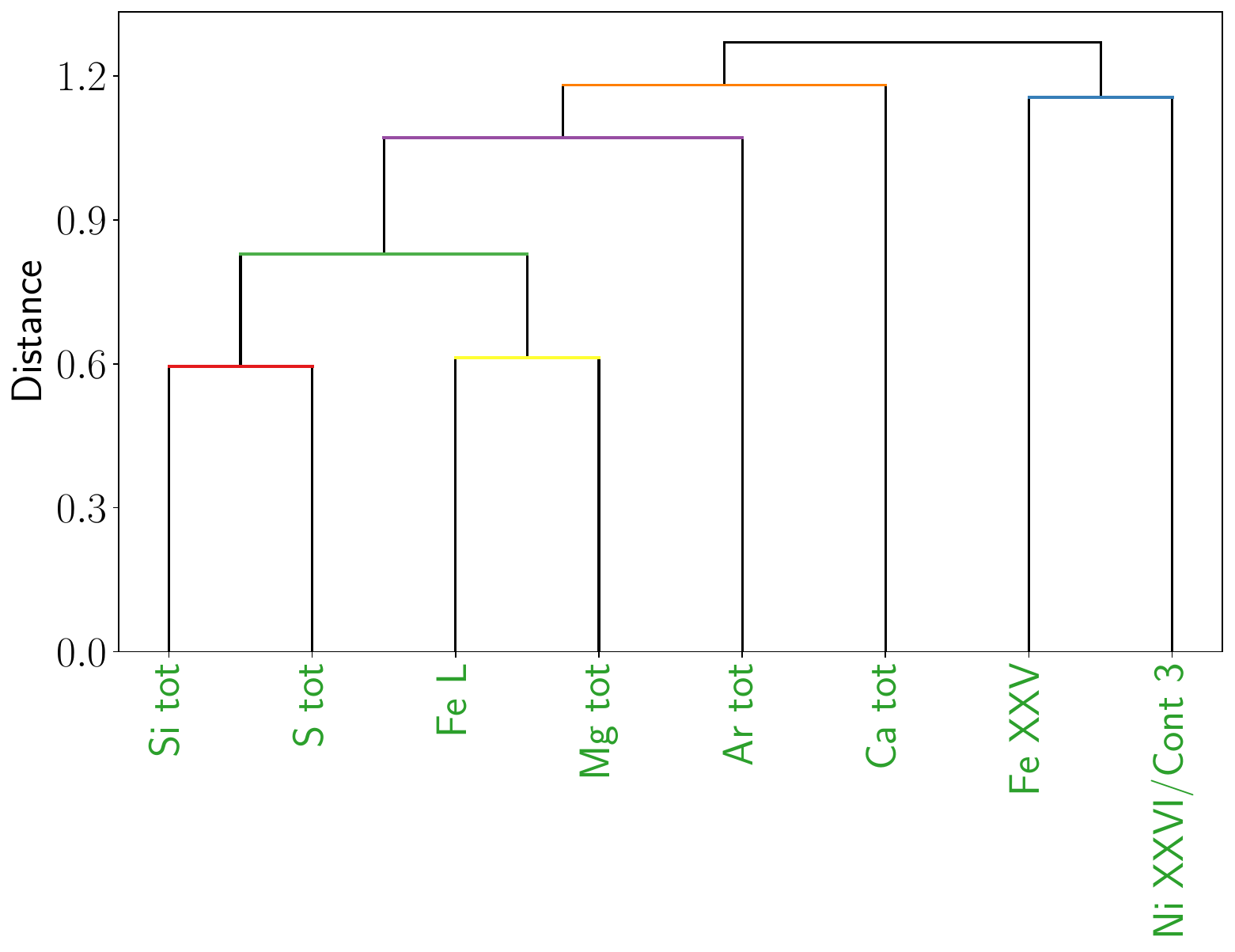}
    \includegraphics[width=0.7\columnwidth]{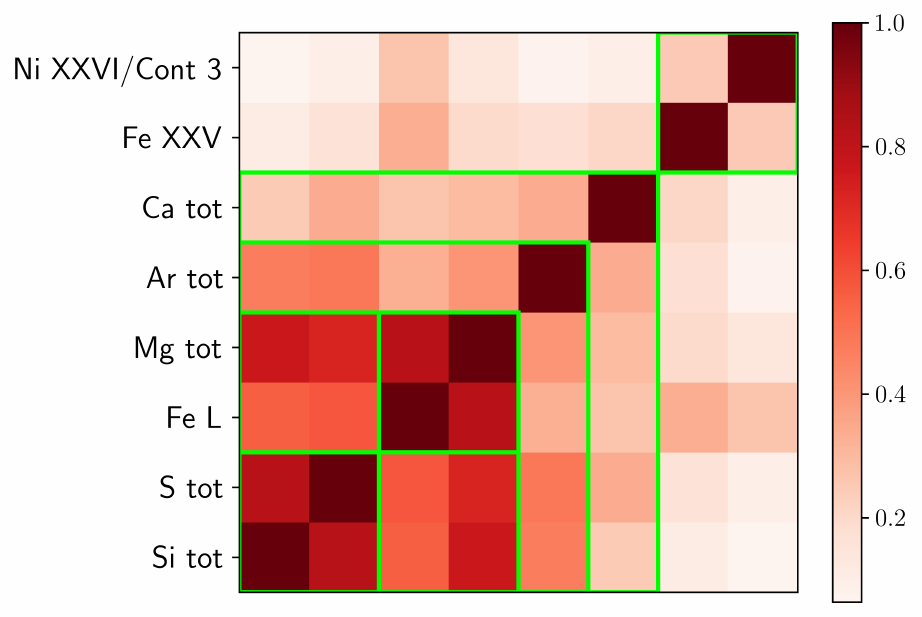}
    \includegraphics[width=0.7\columnwidth]{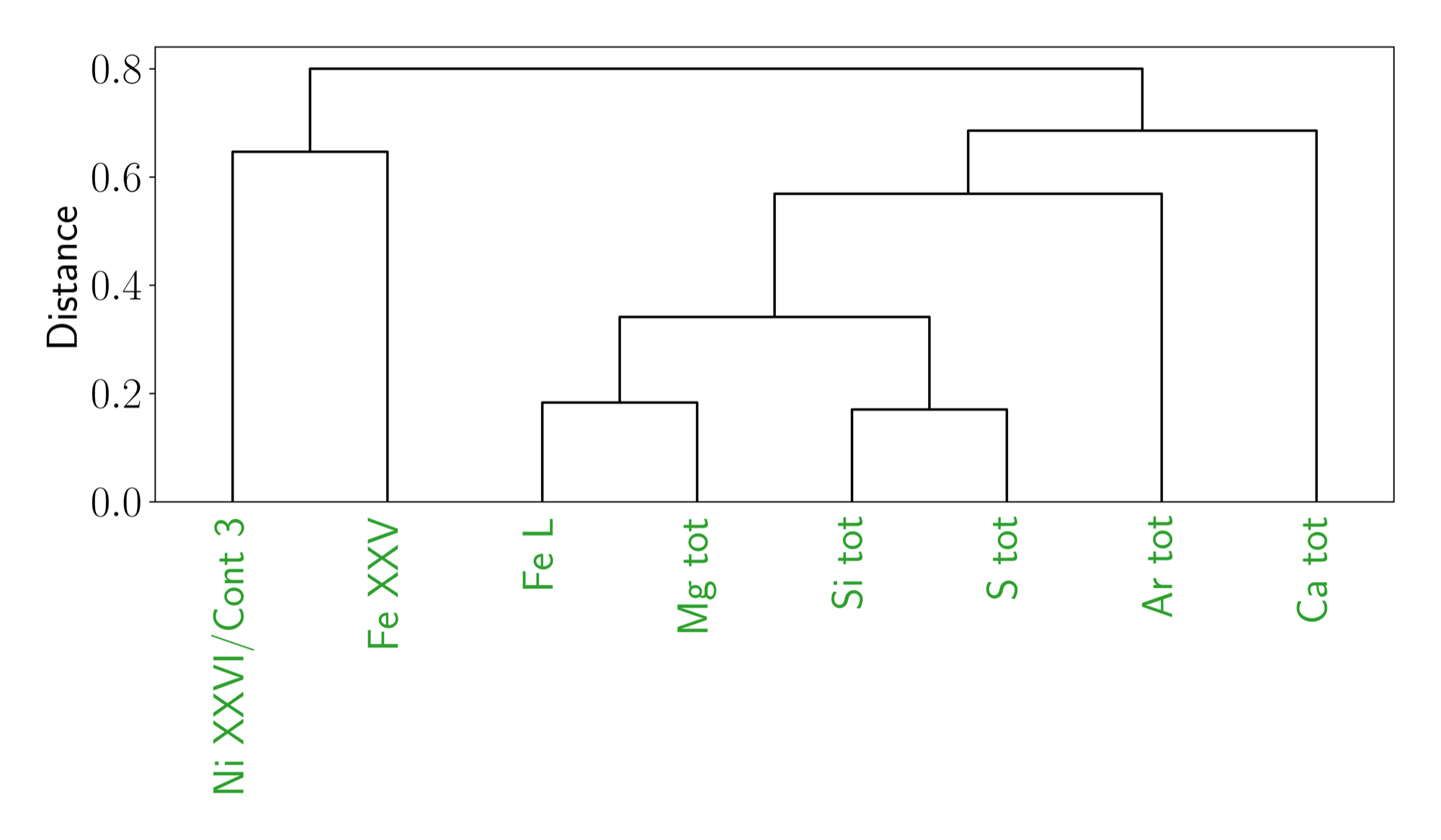}
    \caption{Analysis of the \emph{Chandra} X-ray EW maps of G292.0+1.8.
    The upper panel shows the dendrogram obtained by using the SVHC methods with average linkage and the Frobenius distance of Eq. (\ref{frob}). 
    The central panel shows the distance matrix reordered according to the previous dendrogram.
    The lower panel shows the one obtained by using the average linkage and the cosine distance of Eq. (\ref{cos}).}
    \label{fig:g292ew}
\end{figure}

\begin{figure}[htb] 
    \centering 
    \includegraphics[width=0.7\columnwidth]{{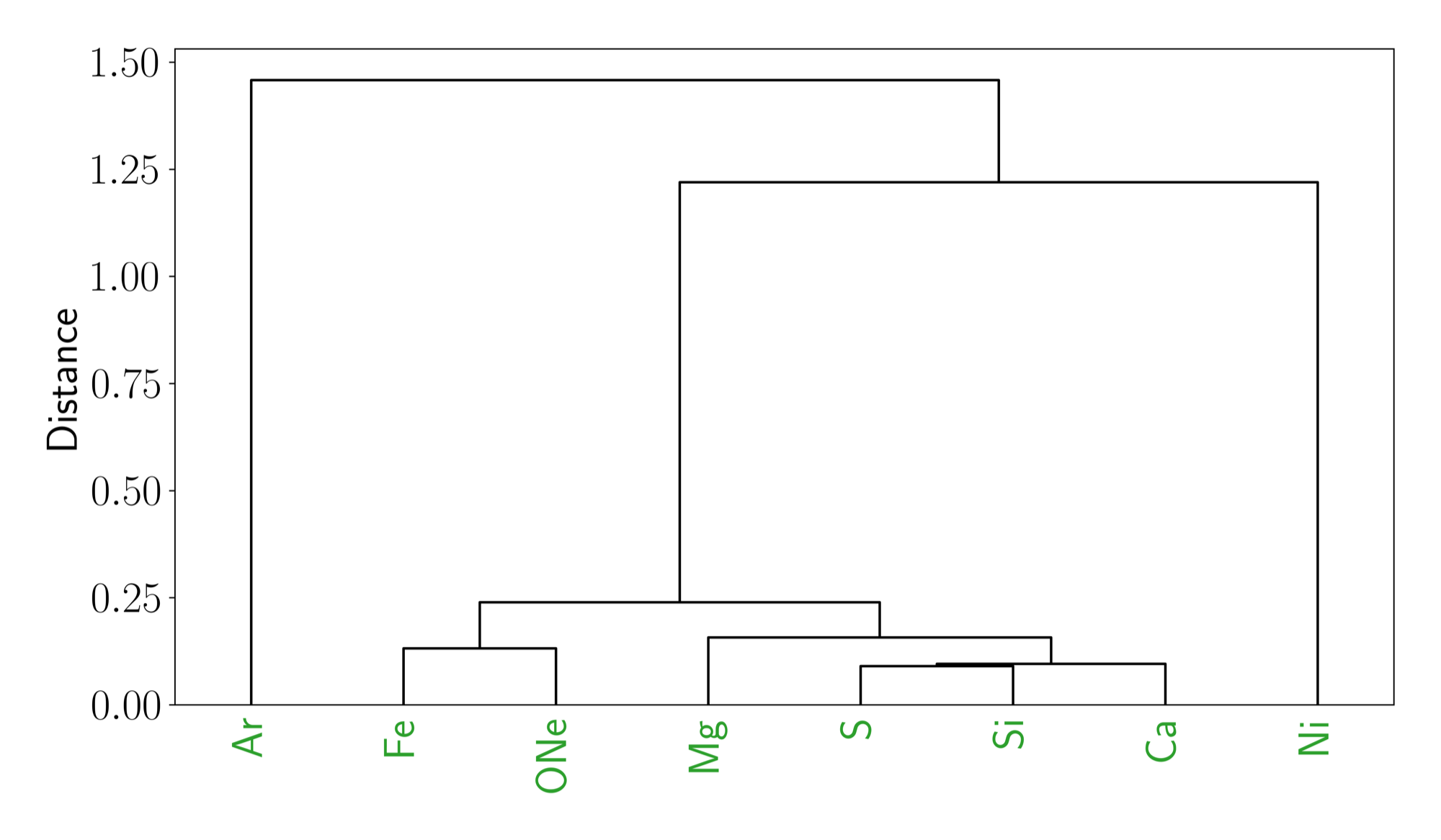}}
    \caption{Analysis of the \emph{Chandra} X-ray EW maps of G292.0+1.8.
             The panel shows the dendrogram obtained by using the GNN embeddings ($k=6$), followed by the application of average linkage with Euclidean distance.}
    \label{fig:g292ewGNN}
\end{figure}

\begin{figure}[htb] 
    \centering 
    \includegraphics[width=0.7\columnwidth]{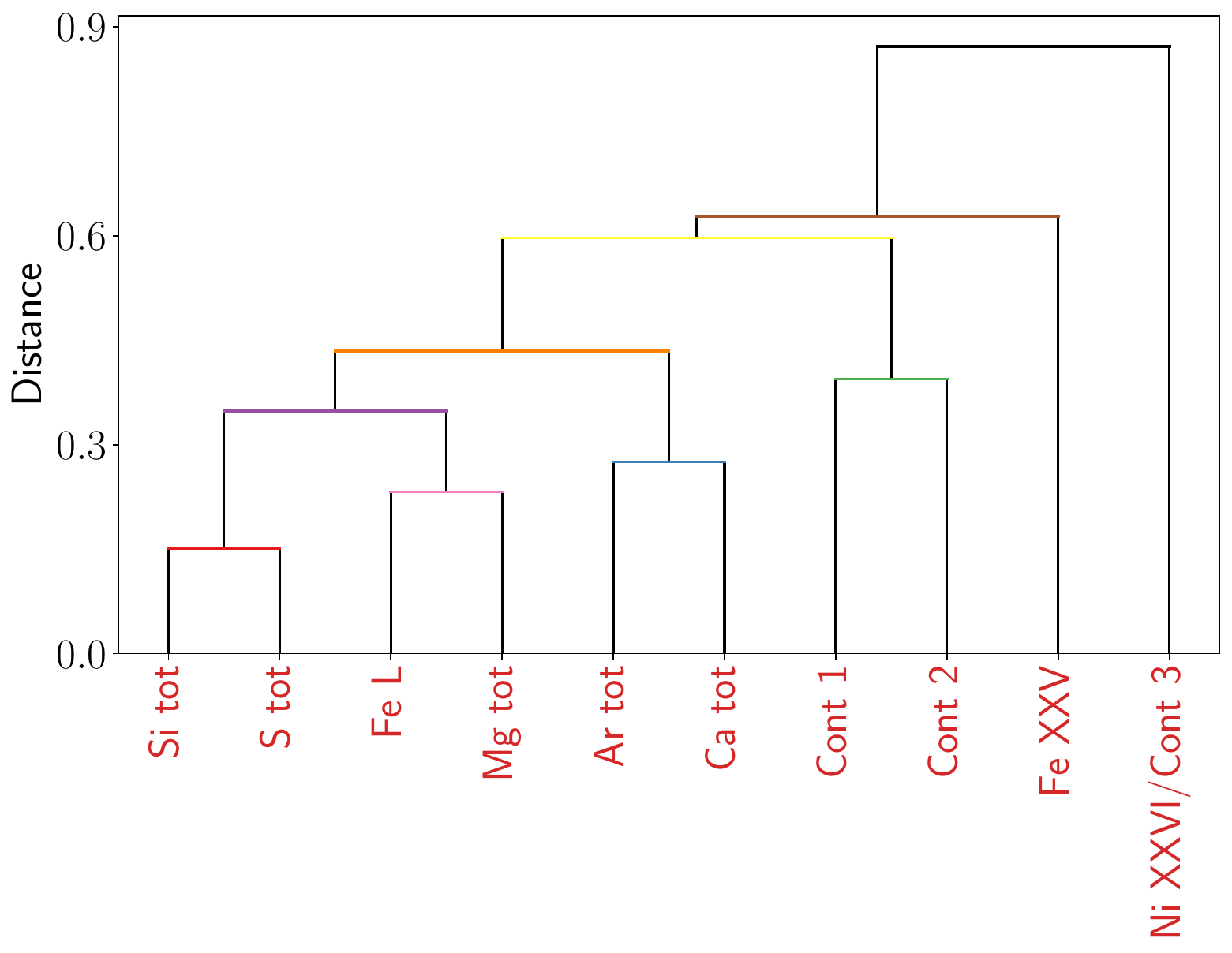}
    \includegraphics[width=0.7\columnwidth]{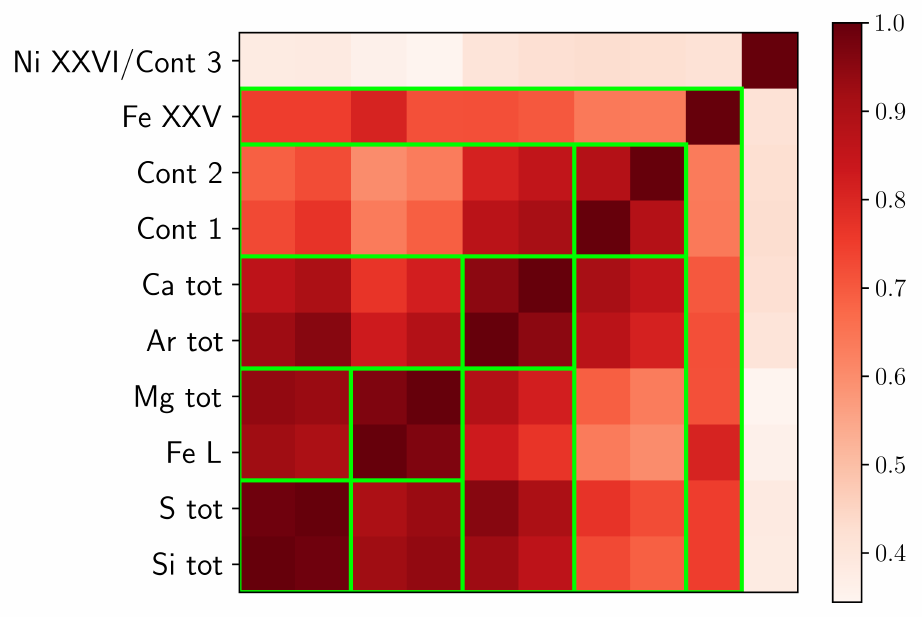}
    \includegraphics[width=0.7\columnwidth]{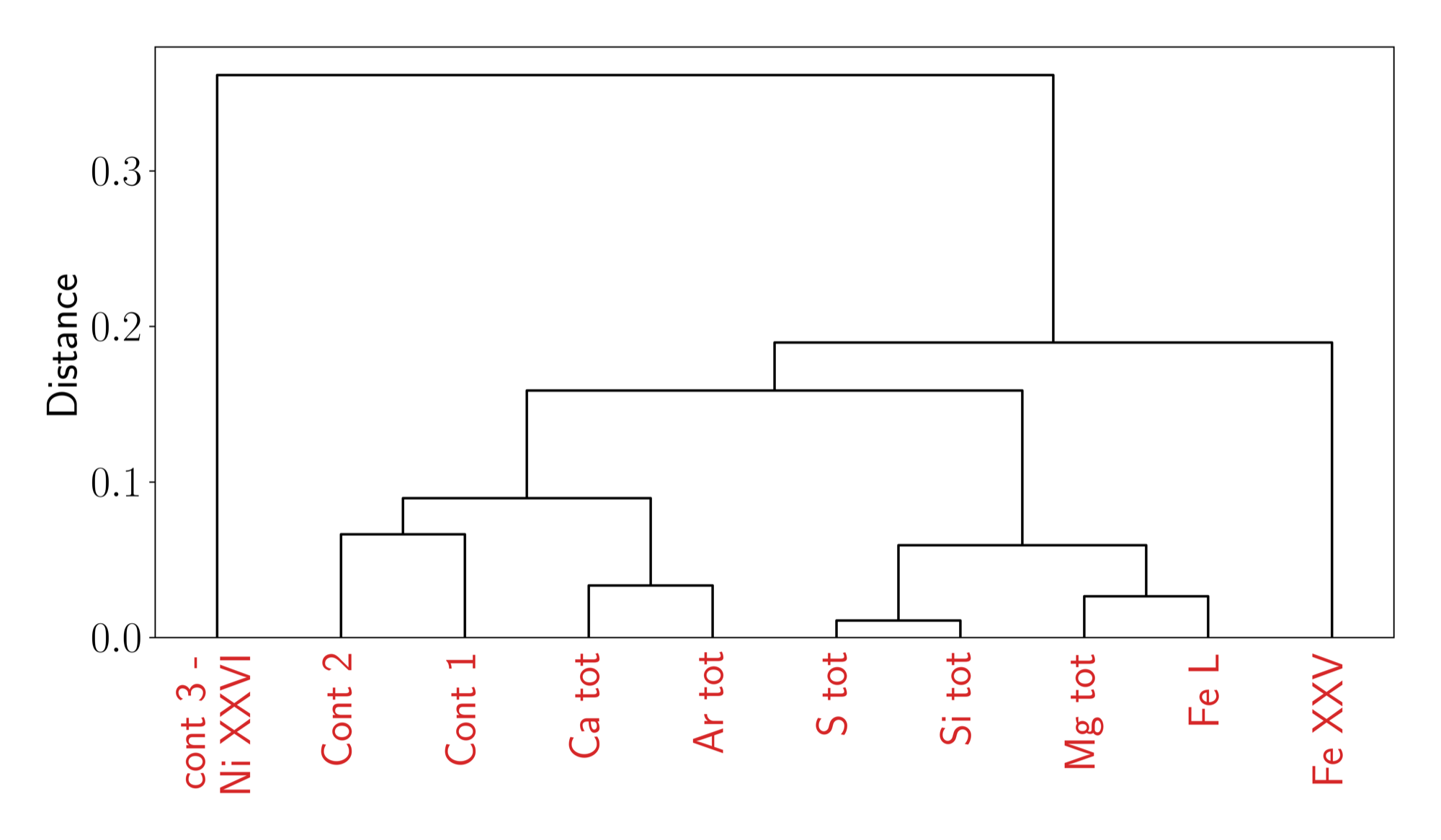}
    \caption{Analysis of the \emph{Chandra} X-ray flux maps of Kepler's SNR.
    The upper panel shows the dendrogram obtained by using the SVHC methods with average linkage and the Frobenius distance of Eq. (\ref{frob}). 
    The central panel shows the distance matrix reordered according to the previous dendrogram.
    The lower panel shows the one obtained by using the average linkage and the cosine distance of Eq. (\ref{cos}).}
    \label{fig:fluxkepler}
\end{figure}

\begin{figure}[htb] 
    \centering 
    \includegraphics[width=0.7\columnwidth]{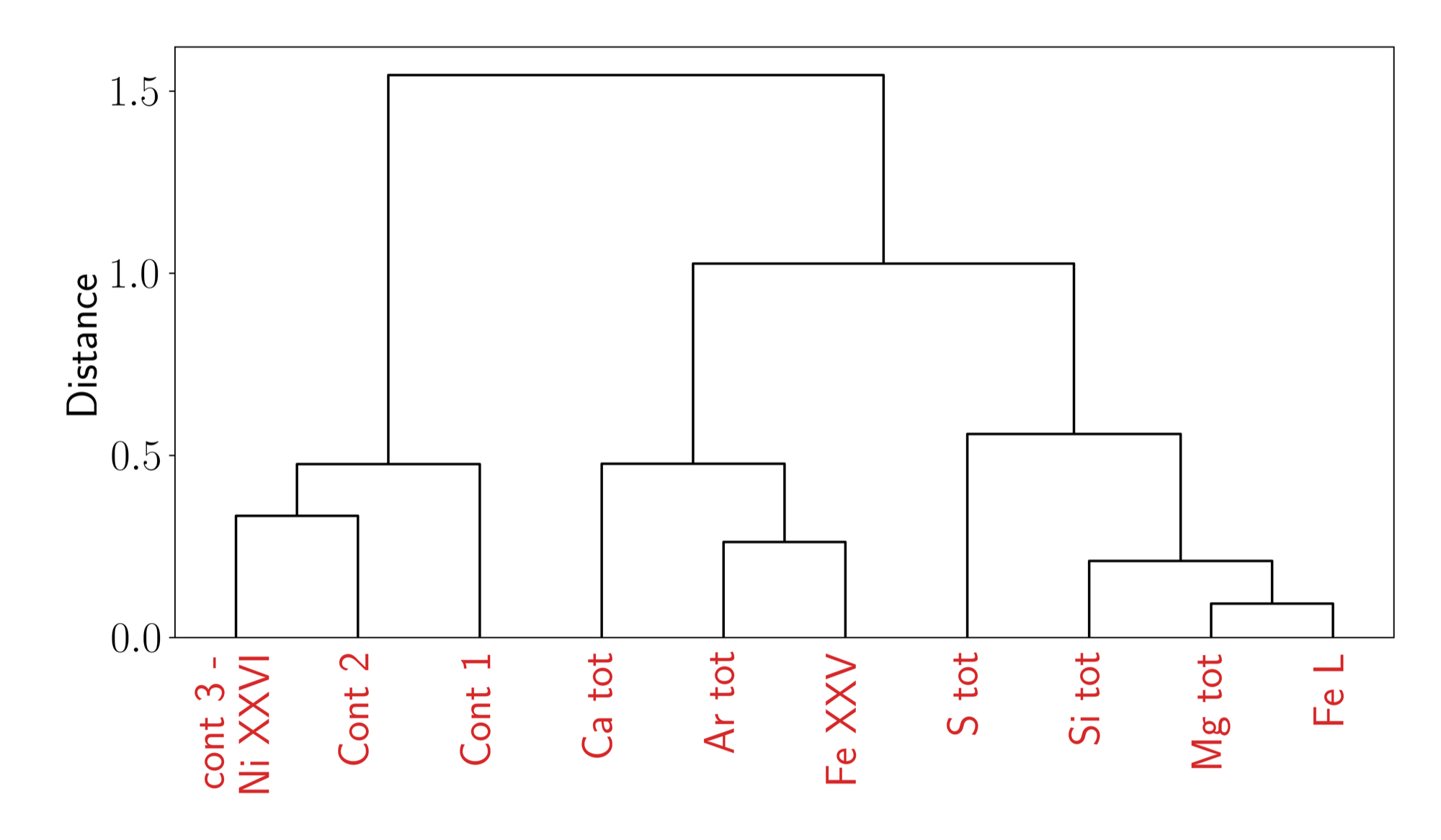}
    \caption{Analysis of the \emph{Chandra} X-ray flux maps of Kepler's SNR.
             The panel shows the dendrogram obtained by using the GNN embeddings ($k=6$), followed by the application of average linkage with Euclidean distance.}
    \label{fig:fluxkeplerGNN}
\end{figure}

\begin{figure}[htb] 
    \centering 
    \includegraphics[width=0.7\columnwidth]{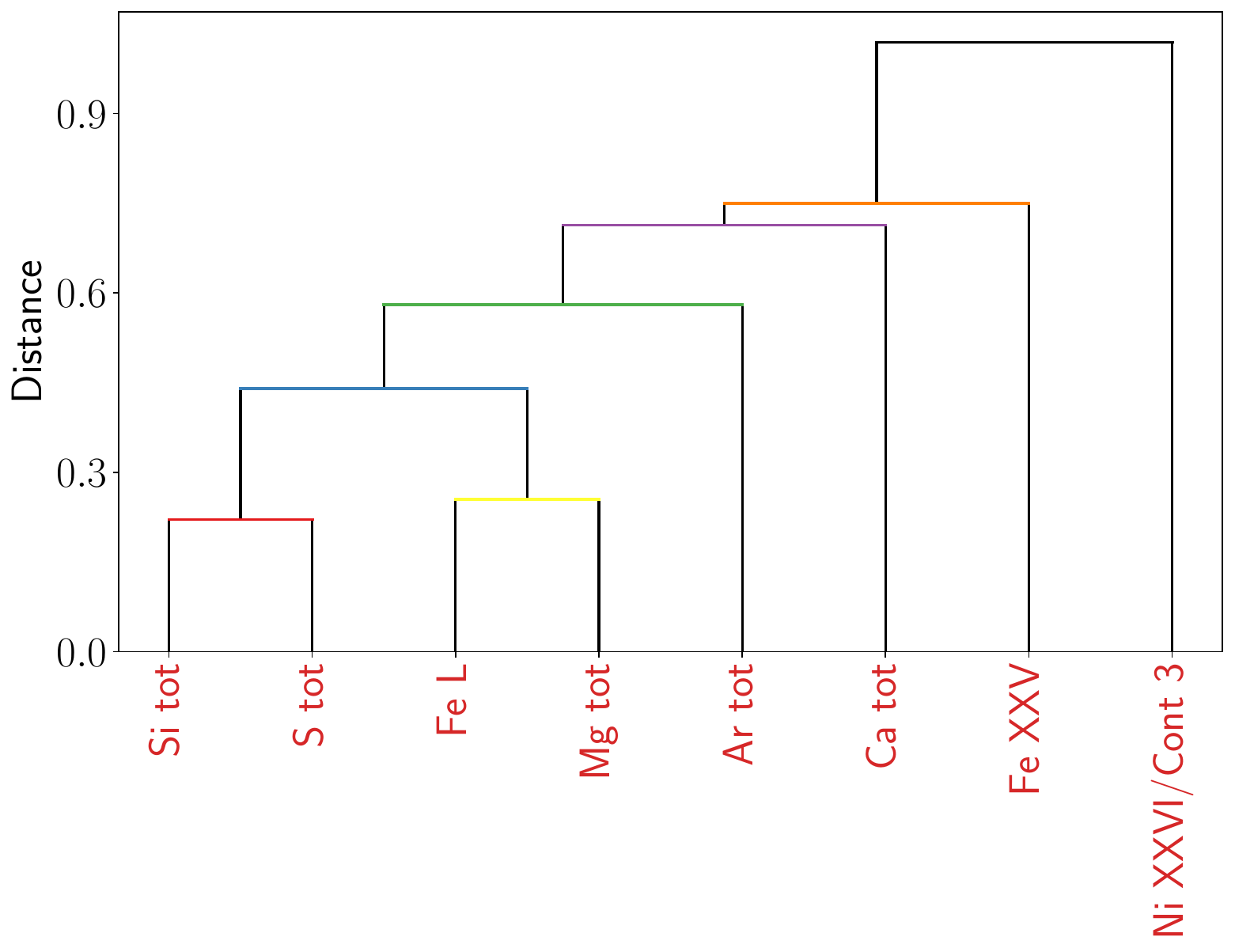}
    \includegraphics[width=0.7\columnwidth]{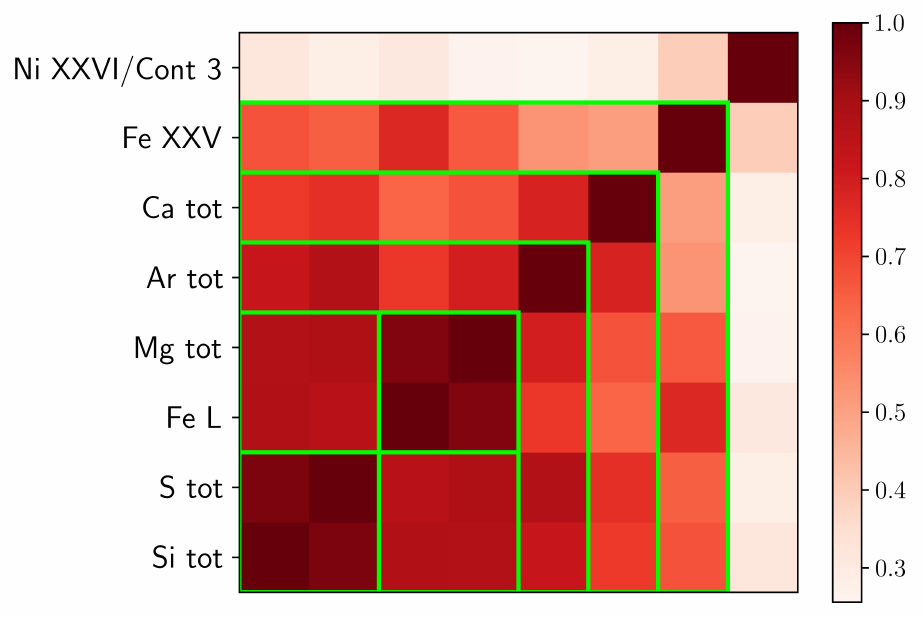}
    \includegraphics[width=0.7\columnwidth]{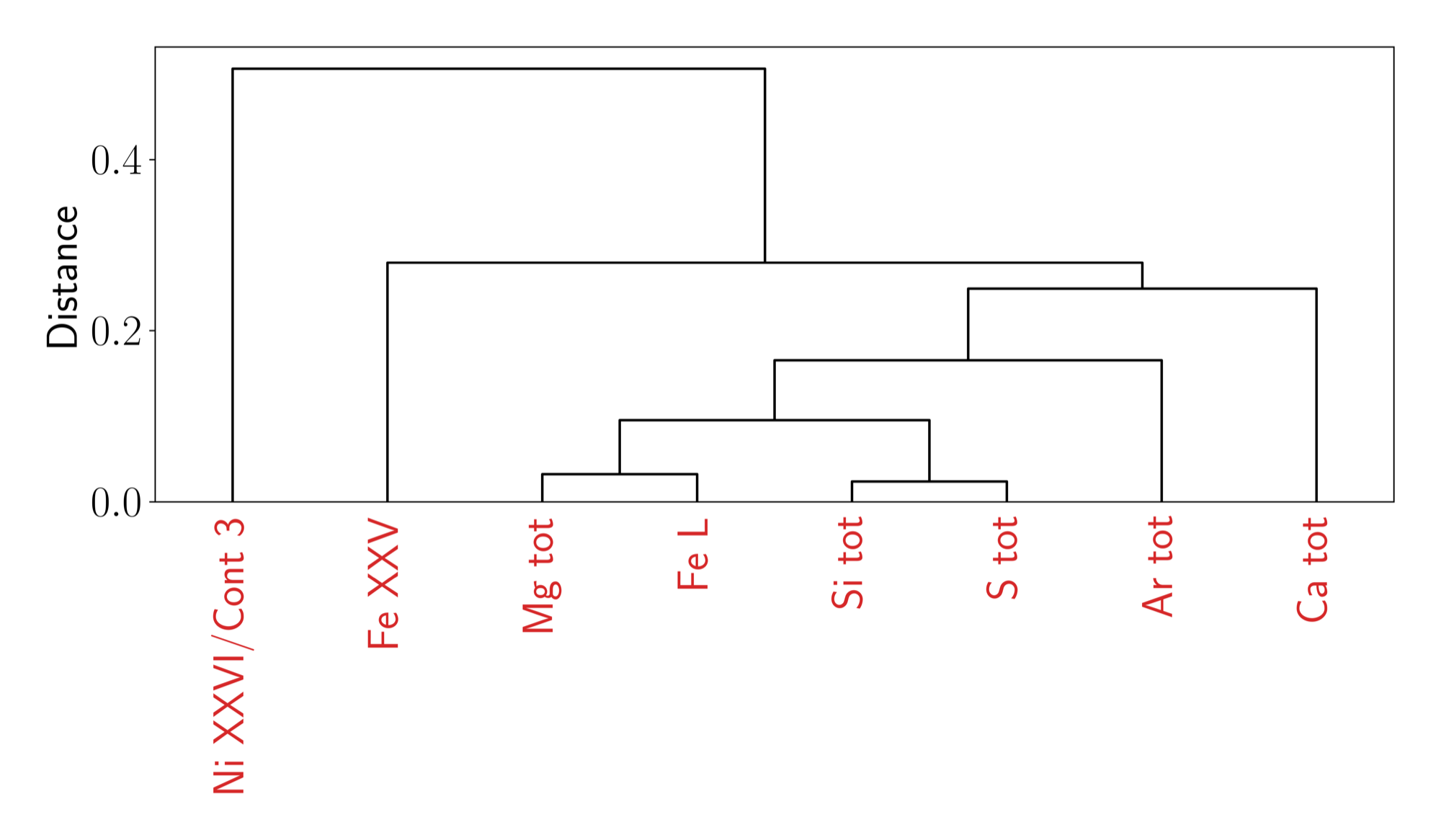}
    \caption{Analysis of the \emph{Chandra} X-ray EW maps of Kepler's SNR.
    The upper panel shows the dendrogram obtained by using the SVHC methods with average linkage and the Frobenius distance of Eq. (\ref{frob}). 
    The central panel shows the distance matrix reordered according to the previous dendrogram.
    The lower panel shows the one obtained by using the average linkage and the cosine distance of Eq. (\ref{cos}).}
    \label{fig:ewkepler}
\end{figure}

\begin{figure}[htb] 
    \centering 
    \includegraphics[width=0.7\columnwidth]{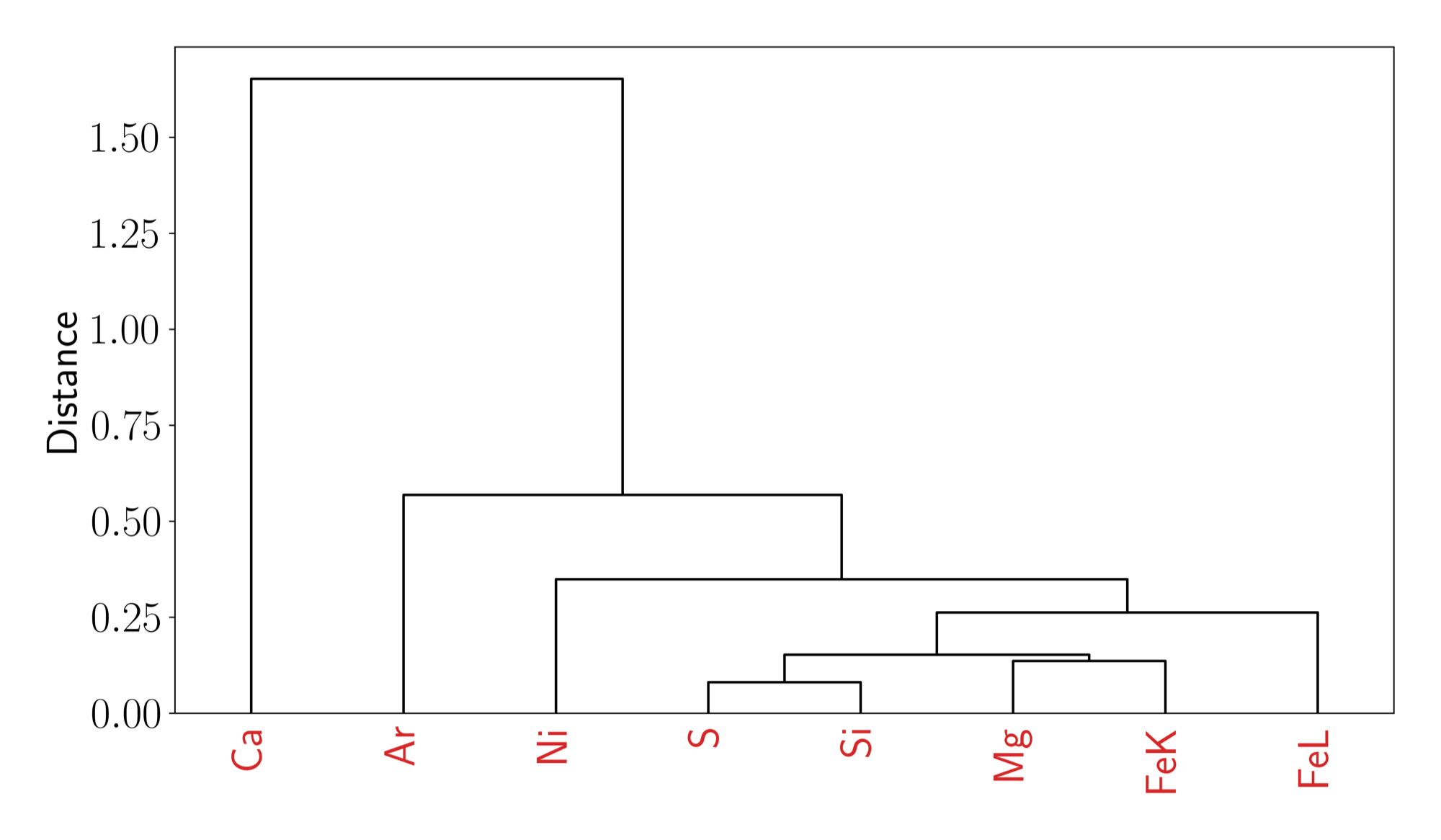}
    \caption{Analysis of the \emph{Chandra} X-ray EW maps of Kepler's SNR.
             The panel shows the dendrogram obtained by using the GNN embeddings ($k=6$), followed by the application of average linkage with Euclidean distance.}
    \label{fig:ewkeplerGNN}
\end{figure}

\begin{figure}[htb] 
    \centering 
    \includegraphics[width=0.7\columnwidth]{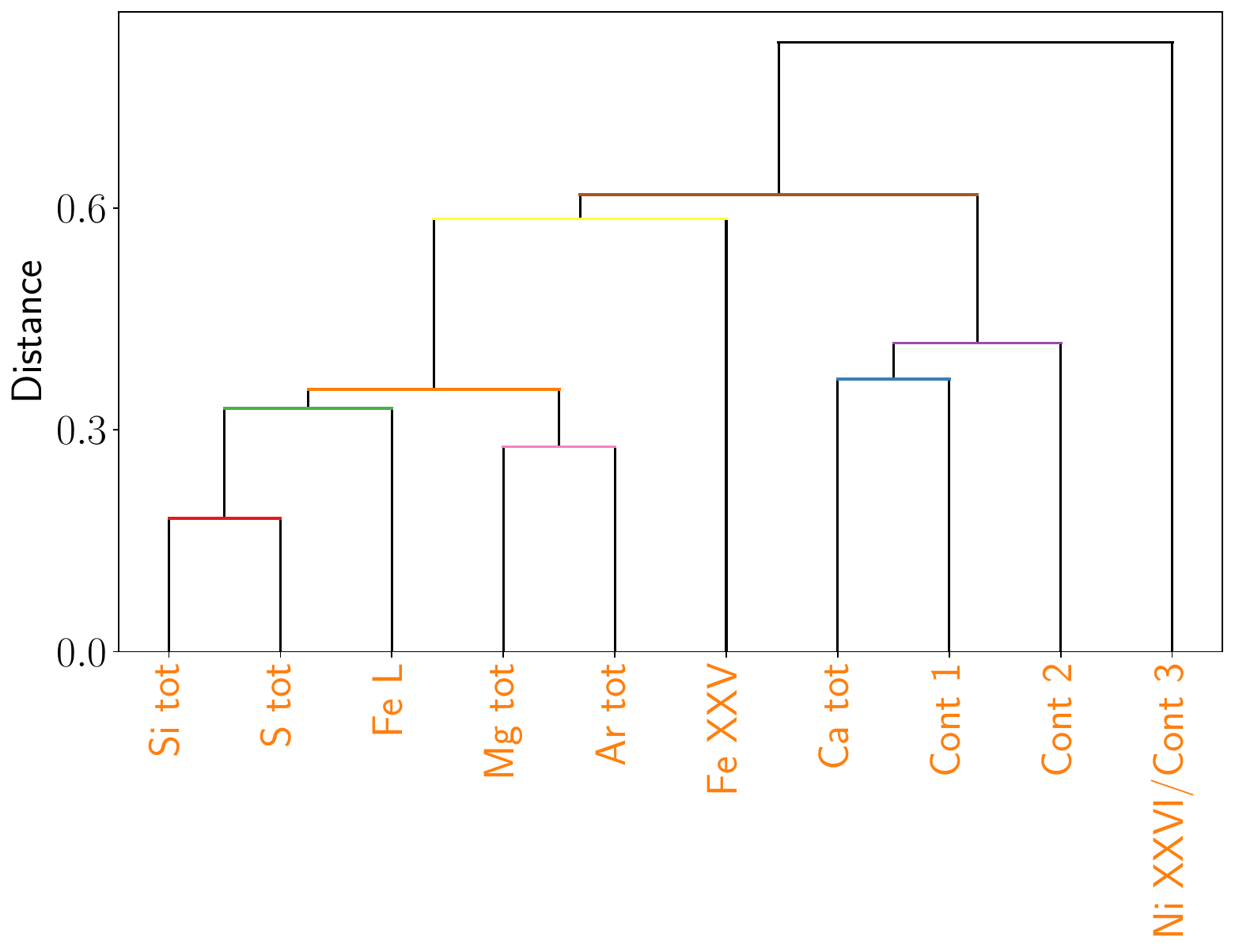}
    \includegraphics[width=0.7\columnwidth]{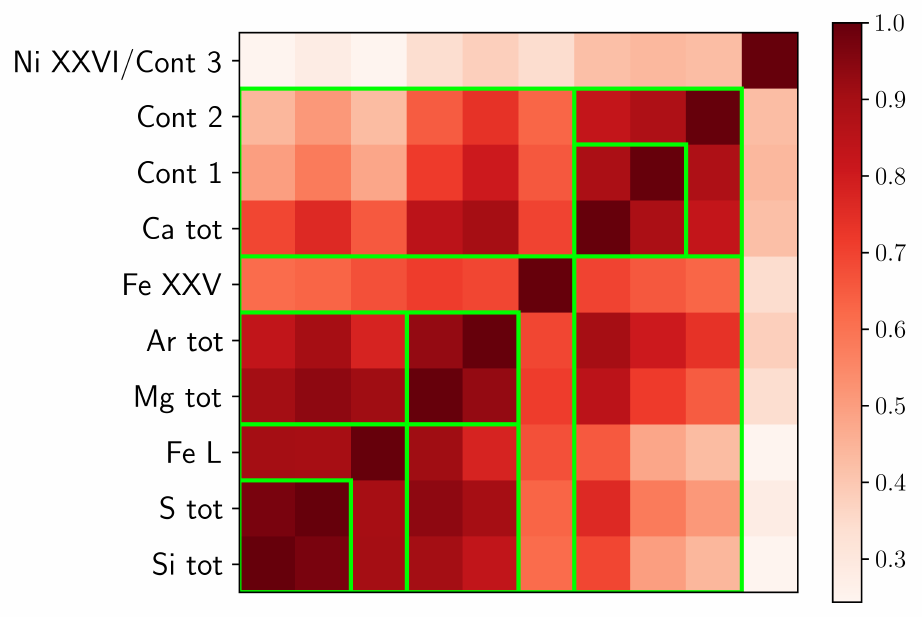}
    \includegraphics[width=0.7\columnwidth]{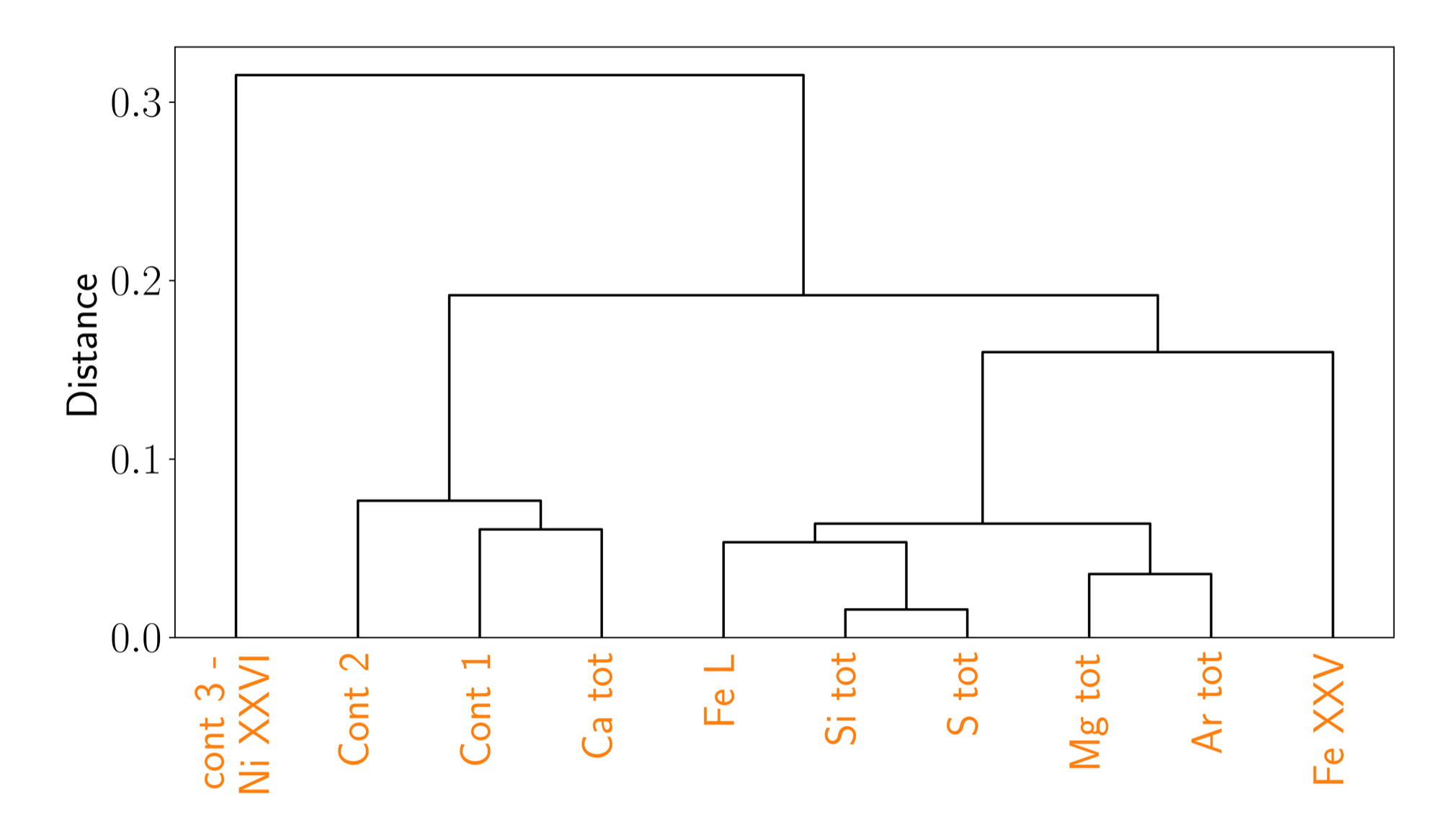}
    \caption{Analysis of the \emph{Chandra} X-ray flux maps of Tycho's SNR.
    The upper panel shows the dendrogram obtained by using the SVHC methods with average linkage and the Frobenius distance of Eq. (\ref{frob}). 
    The central panel shows the distance matrix reordered according to the previous dendrogram.
    The lower panel shows the one obtained by using the average linkage and the cosine distance of Eq. (\ref{cos}).}
    \label{fig:tychoflux}
\end{figure}

\begin{figure}[htb] 
    \centering 
    \includegraphics[width=0.7\columnwidth]{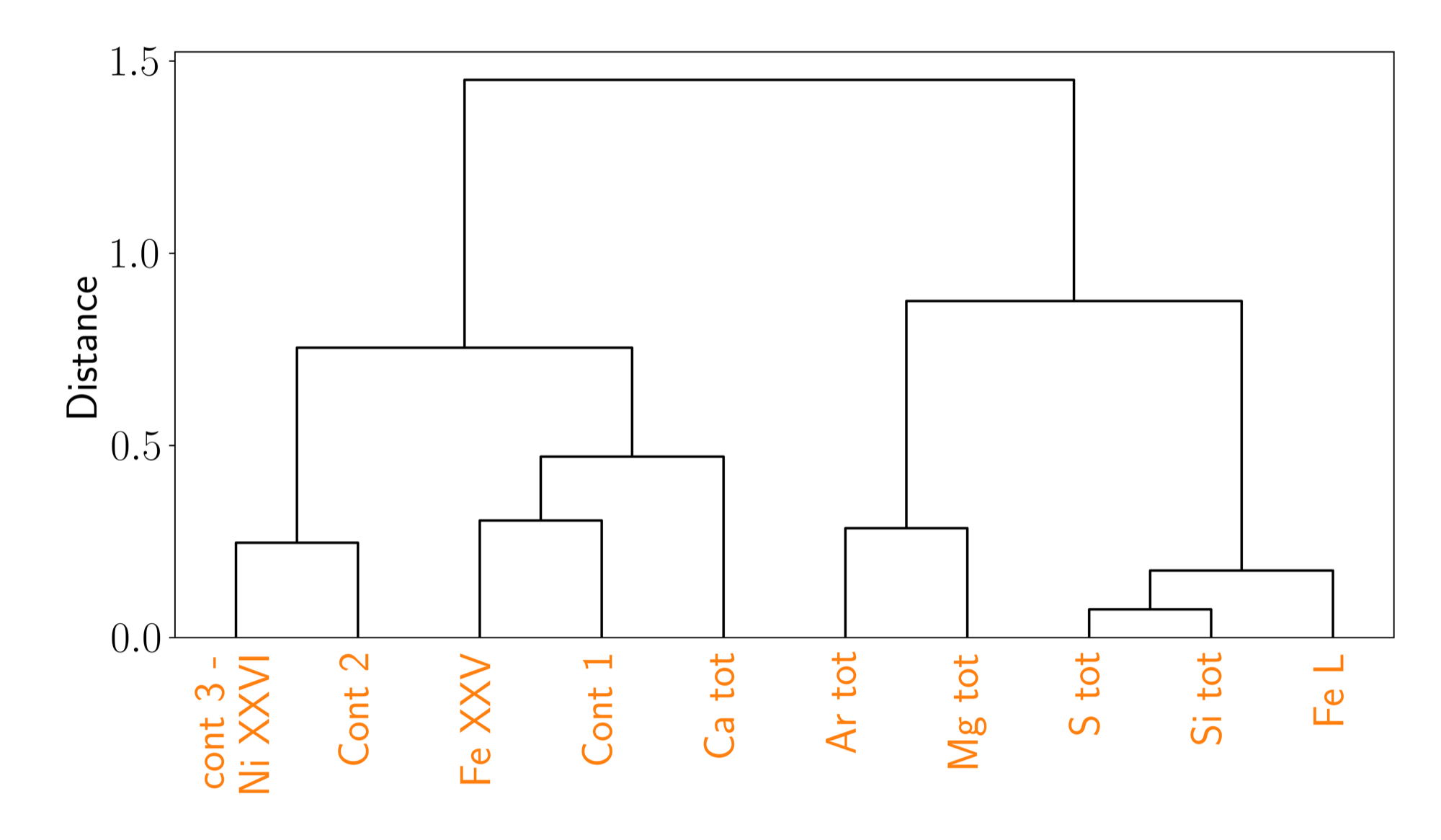}
    \caption{Analysis of the \emph{Chandra} X-ray flux maps of Tycho's SNR.
             The panel shows the dendrogram obtained by using the GNN embeddings ($k=6$), followed by the application of average linkage with Euclidean distance.}
    \label{fig:tychofluxGNN}
\end{figure}

\begin{figure}[htb] 
    \centering 
    \includegraphics[width=0.7\columnwidth]{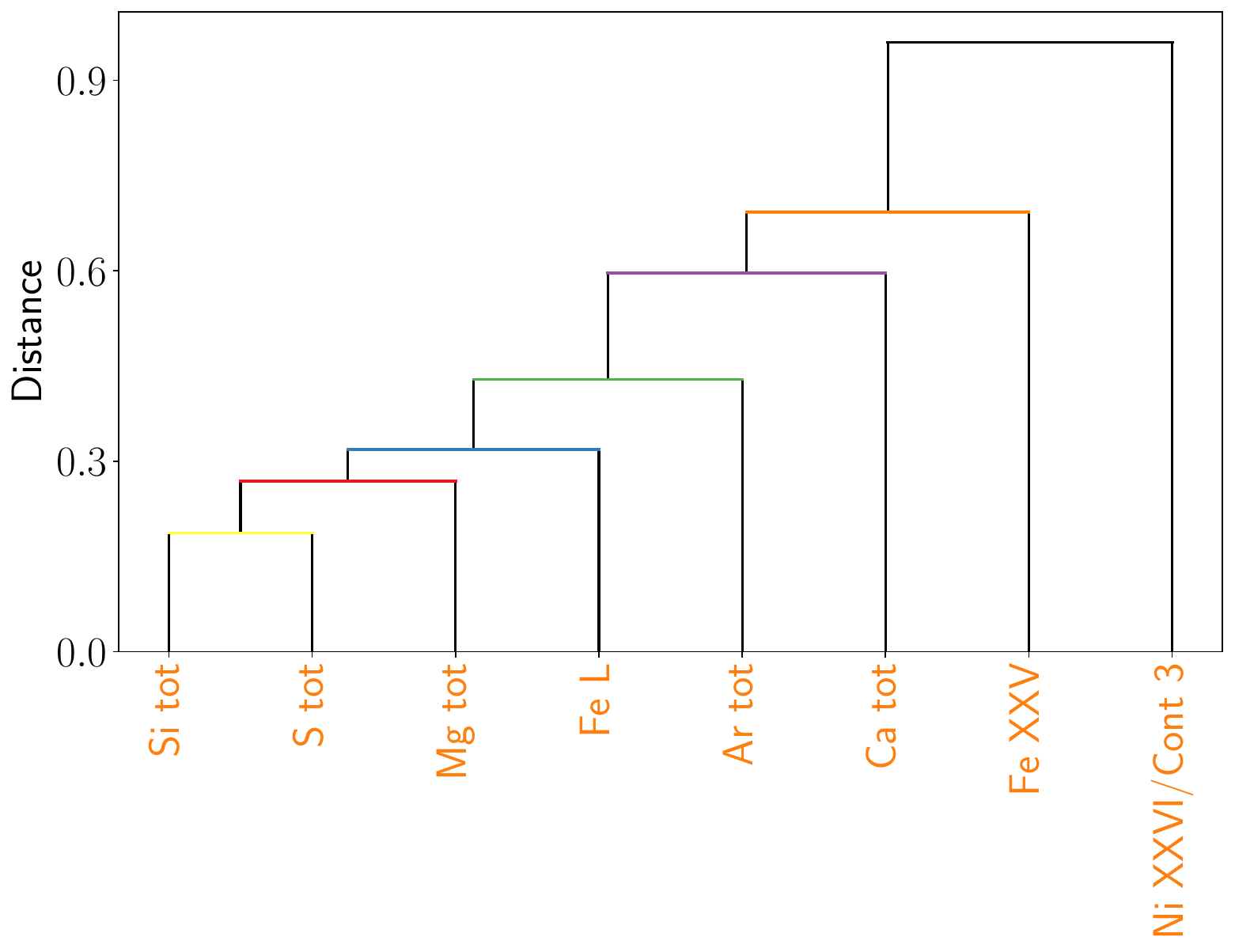}
    \includegraphics[width=0.7\columnwidth]{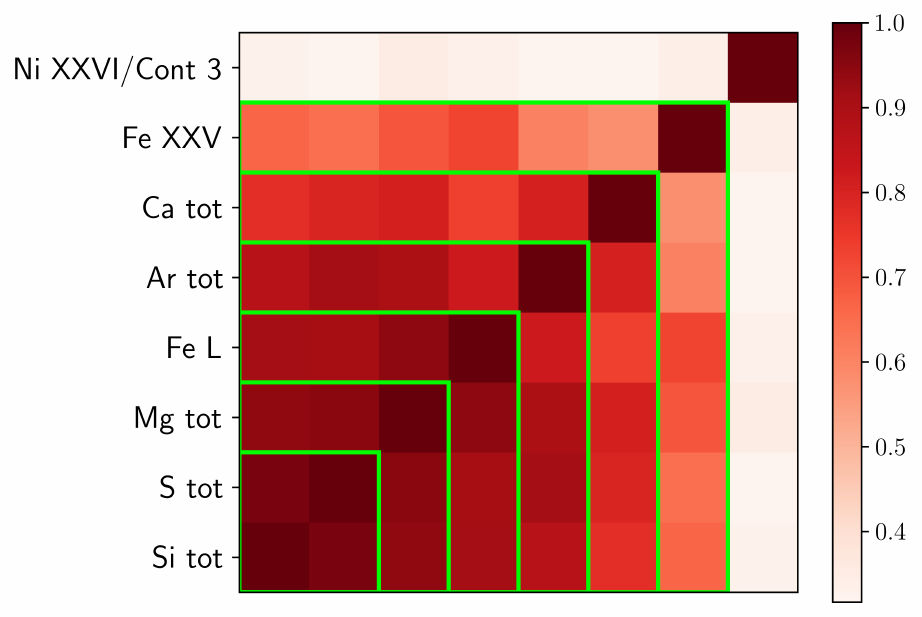}
    \includegraphics[width=0.7\columnwidth]{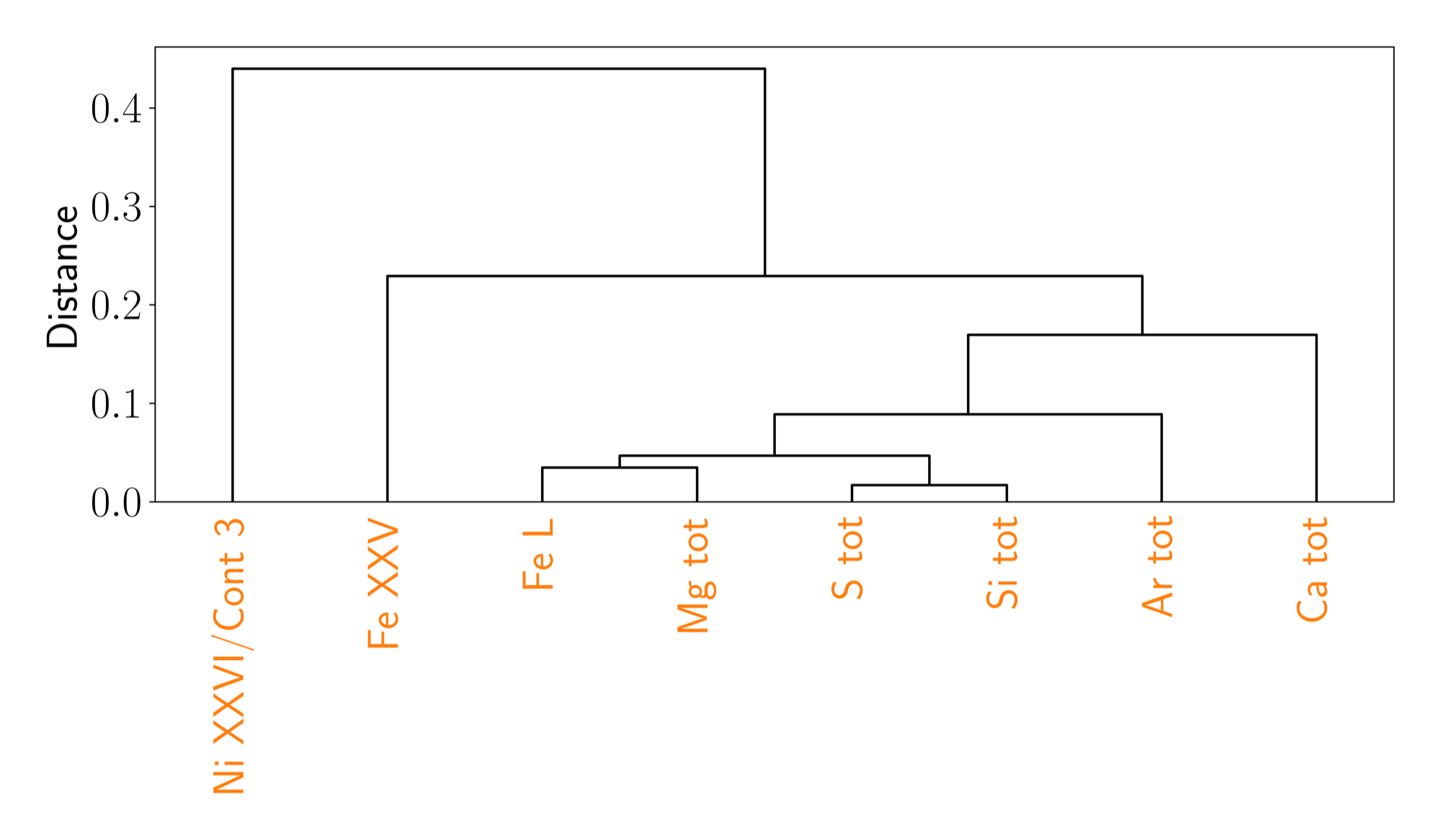}
    \caption{Analysis of the \emph{Chandra} X-ray EW maps of Tycho's SNR.
    The upper panel shows the dendrogram obtained by using the SVHC methods with average linkage and the Frobenius distance of Eq. (\ref{frob}). 
    The central panel shows the distance matrix reordered according to the previous dendrogram.
    The lower panel shows the one obtained by using the average linkage and the cosine distance of Eq. (\ref{cos}).}
    \label{fig:tychoew}
\end{figure}

\begin{figure}[htb] 
    \centering 
    \includegraphics[width=0.7\columnwidth]{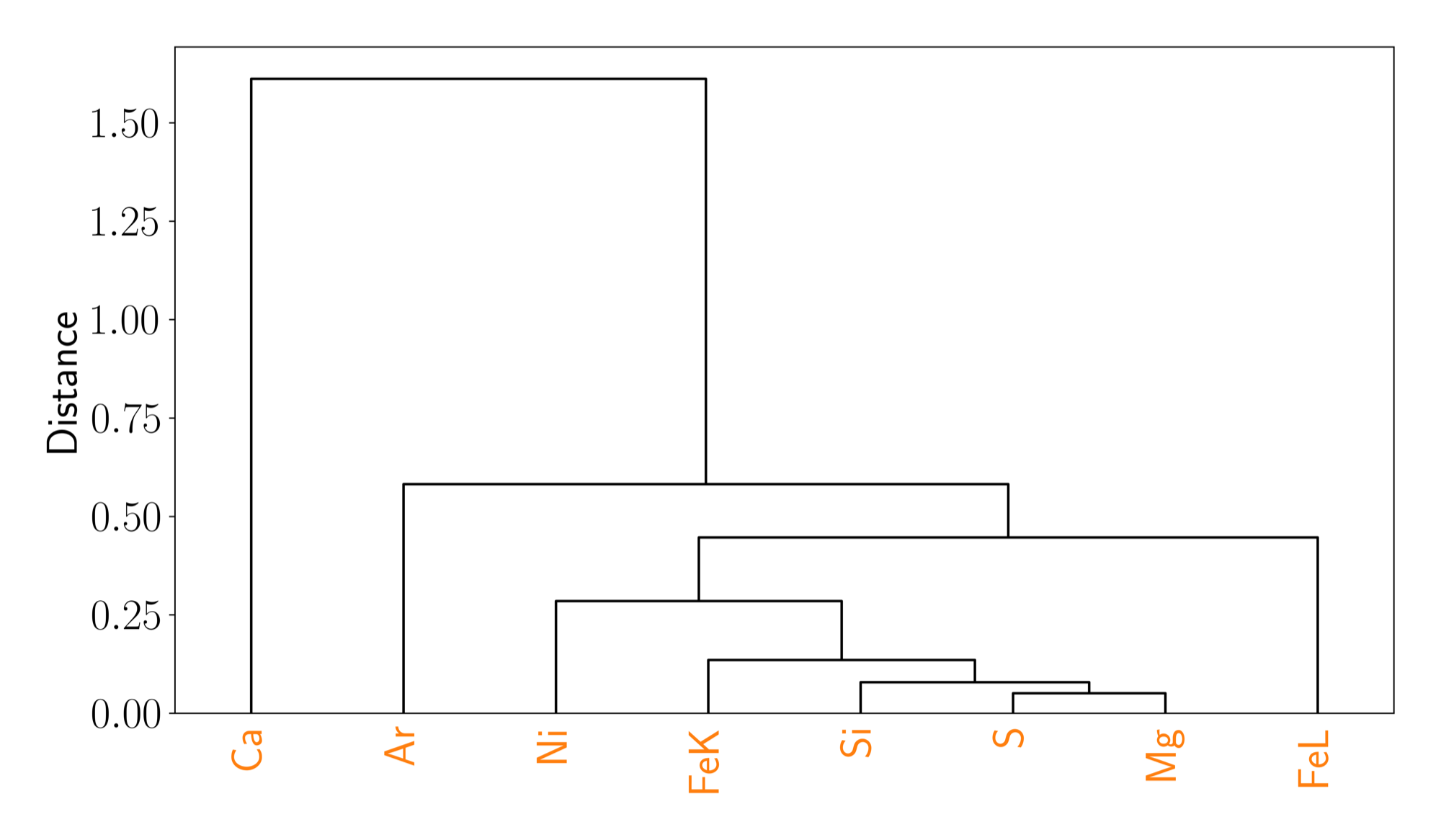}
    \caption{Analysis of the \emph{Chandra} X-ray EW maps of Tycho's SNR.
             The panel shows the dendrogram obtained by using the GNN embeddings ($k=6$), followed by the application of average linkage with Euclidean distance.}
    \label{fig:tychoewGNN}
\end{figure}

\end{appendix}

\end{document}